\documentclass[aps, twocolumn, superscriptaddress, pra]{revtex4-2}
\usepackage{ORI_Group_style}

\newcommand{\ww}{\bold{w}}

\newcommand{\veps}{\varepsilon}
\def\dpp{\boldsymbol{\wp}}
\newcommand{\OD}{\text{OD}}

\newcommand{\cop}{\hat{c}}
\newcommand{\cdop}{\hat{c}^\dag}
\newcommand{\rhop}{\hat{\rho}}

\newcommand{\SMreflist}{Clemens2003, Penrose2003,jackson,Bellando2021,Mok2026,%
Koltchinskii2000,Williams2000,Ressayre1976,Ressayre1977,Gradshteyn7th,%
Tricomi1950,Janaswamy2020,Svidzinsky2008,Shahmoon2017,%
Perczel2017,Zhang2019,AsenjoGarcia2017PRX,vonMilczewski2025,Abella1966,%
diamond2016cvxpy,AsenjoGarcia2017PRA,Cardenas-Lopez2023,Zhang2026}

\begin{document}

\title{Optical depth dictates universal bounds on many-body decay in atomic ensembles}

\author{Cosimo C. Rusconi}
\affiliation{Instituto de F\'isica Fundamental - Consejo Superior de Investigaciones Cient\'ifica (CSIC), Madrid, Espa\~na.}
\affiliation{Department of Physics, Columbia University, New York, New York 10027, USA.}
\author{Eric Sierra}
\affiliation{Department of Physics, Columbia University, New York, New York 10027, USA.}
\author{Wai-Keong Mok}
\affiliation{Institute for Quantum Information and Matter, California Institute of Technology, Pasadena, CA 91125, USA.}
\author{Avishi Poddar}
\affiliation{Department of Physics, Harvard University, Cambridge, Massachusetts 02138, USA.}
\author{Simon B. J\"ager}
\affiliation{Physikalisches Institut, University of Bonn, Nussallee 12, 53115 Bonn, Germany.}
\author{Ana Asenjo-Garcia}
\affiliation{Department of Physics, Columbia University, New York, New York 10027, USA.}

\date{\today}

\begin{abstract}
Cooperative emission is well understood for idealized symmetric systems, but its fundamental limits in spatially extended, free-space ensembles remain an open question. Here, we derive a universal law for the scaling of the maximum photon emission rate with system size that unifies both ordered arrays and disordered atomic clouds in arbitrary dimensions at fixed density. We demonstrate that, for a fixed atomic density, the maximum emission rate scales universally as the product of the atom number and the system's geometric optical depth. The geometric optical depth differs from the result of the Beer-Lambert law at high density and encodes the dimensional scaling across all regimes from independent emission to the Dicke limit. Furthermore, we establish a scaling law for directional detection, revealing that the observed rate depends on the detector's numerical aperture: small apertures yield Dicke-like quadratic scaling, whereas large apertures recover our integrated universal bound. Our results establish optical depth as the parameter governing many-body cooperative emission in both ordered and disordered ensembles, and reveal that directional and total-emission scalings must be carefully distinguished in experimental settings.
\end{abstract}

\maketitle
Understanding how ensembles of excited atoms interact with light is a central question in physics. Its answer underpins phenomena ranging from new light sources~\cite{Meiser2009, Bohnet2012, Bychek2025} to driven-dissipative phase transitions~\cite{Carmichael1980, Ferioli2023, Agarwal2024, Goncalves2024, Ostermann2024, Ruostekoski2024}. Despite its central importance, an exact solution is known only for two special cases. Atoms separated by distances much larger than the optical wavelength emit independently, and the total emission rate decreases monotonically from an initial value that scales with the atom number $N$. In contrast, atoms confined well within an optical wavelength (or positioned at the antinodes of a single-mode cavity) couple identically to the electromagnetic field, giving rise to exact permutational symmetry and enabling an exact solution~\cite{Dicke1954, Lee1977a, Lee1977b, Rupasov1984, Holzinger2025arXiv}. The collective coupling to the field enhances the maximum emission rate up to a value scaling as  $N^2$~\cite{Dicke1954}. Beyond these limits---and, in particular, for extended ensembles in free space--- the reduced symmetry prevents an exact characterization of the many-body radiative dynamics. One must rely instead on approximate numerical methods~\cite{Clemens2004, RubiesBigorda2023, Mink2023} or heuristic effective models~\cite{Bonifacio1970, Bonifacio1971, Rehler1971, Ressayre1976, Ressayre1977, Holzinger2025c, Holzinger2025b}, which cannot be validated due to lack of exact results.

While solving the full many-body radiative dynamics remains intractable, substantial progress can be made by focusing on a simpler but physically central question: \textit{what is the maximum photon emission rate an ensemble can sustain?} This quantity has been mapped to the ground-state problem of an effective spin Hamiltonian~\cite{Mok2023}, enabling rigorous lower and upper bounds for general emitter configurations~\cite{Mok2026} through tools developed for approximating ground-state energies~\cite{barthel2012solving, baumgratz2012lower, Bravyi2019, baccari2020verifying, parekh2021application, Brandao2013}. For ordered atomic arrays~\cite{Endres2016, Barredo2016, Kim2016, Kumar2018, kaufman, Rui2020, Srakaew2023}, these bounds are asymptotically tight and reveal a universal scaling law for the maximum emission rate that depends only on the lattice dimensionality. However, this result relies crucially on spatial order, and its validity for realistic disordered systems remains unclear. Moreover, the physical mechanism underlying the scaling -- and, in particular, whether it is tied to crystalline order or reflects a more general property -- has not been elucidated.

\begin{figure}[t!]
	\includegraphics[width=\columnwidth]{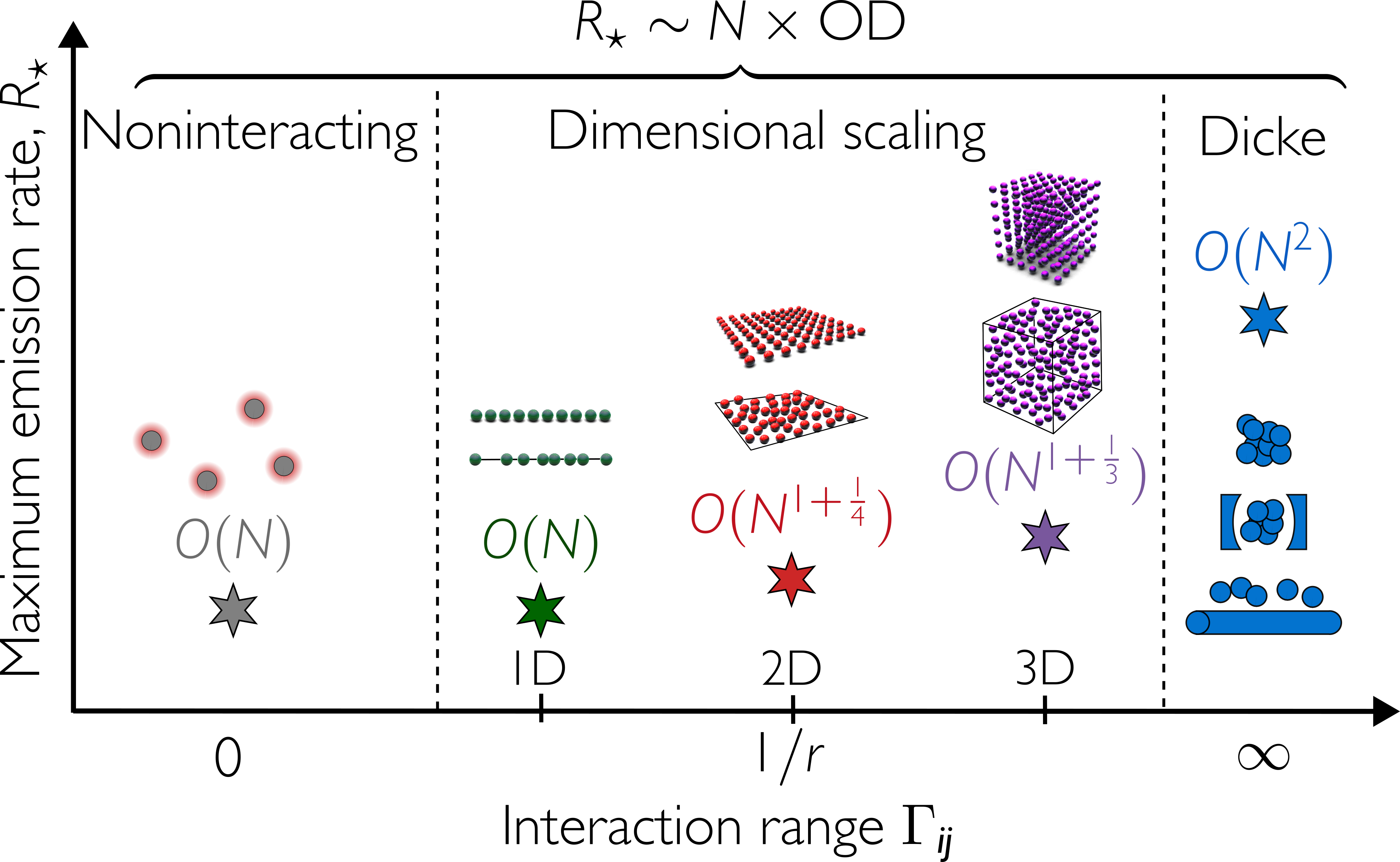}
    \caption{The ensemble's optical depth (OD) sets the scaling with system size $N$ of the maximum emission rate $R_\star$ across all regimes, identified by the longest interaction range of the \emph{dissipative interactions} $\Gamma_{ij}$. The Dicke-like scaling applies to atoms at a point, in cavities, or along waveguides.}\label{fig:Fig1}
\end{figure}

In this Letter, we establish a direct connection between the maximum emission rate, $R_\star$, of a generic atomic ensemble and its geometric optical depth. Specifically, we show that
\be\label{eq:Scaling_Law_OD}
    R_\star \sim \Gamma_0\, N \times \text{OD},
\ee
where $N$ is the atom number, $\Gamma_0$ is the single-atom decay rate, and OD is the geometric optical depth ($\text{OD}\sim N\Delta\Omega$, with $\Delta\Omega$ the solid angle into which the ensemble radiates). We remark that the geometric OD does not contain the effects of Hamiltonian interactions, and thus differs from the result of the Beer-Lambert law at high densities~\cite{Andreoli2021}. The scaling law in Eq.~\eqref{eq:Scaling_Law_OD} unifies regimes from single-particle to superradiant emission [Fig.~\ref{fig:Fig1}]. In the non-interacting limit ($d \gg \lambda_0$), $\mathrm{OD} \to O(1)$ and the scaling reduces to $R_\star \sim N\Gamma_0$. For dense extended ensembles of dimensionality $D$, the OD scales as $N^{\frac{1}{2} - \frac{1}{2D}}$, yielding a dimensional scaling for $R_\star$.
In the Dicke limit ($L \ll \lambda_0$), $\mathrm{OD} \sim N$ recovers $R_\star \sim N^2 \Gamma_0$. More generally, \eqnref{eq:Scaling_Law_OD} holds across electromagnetic environments.
For atoms coupled to a single-mode cavity or a waveguide, $\mathrm{OD} \sim N$ and $R_\star\sim N^2\Gamma_\text{1D}$, where  $\Gamma_\text{1D}$ replaces $\Gamma_0$ as the single-atom emission rate into the privileged mode. The OD, which in the linear optics regime sets the strength of light-matter coupling~\cite{Guerin2017, Chang2018}, thereby acquires a fundamental role also in the many-body problem. Equation~\eqref{eq:Scaling_Law_OD} governs the total, angle-integrated emission.

We further derive a scaling law for directional detection, showing that the maximum detected intensity depends on the numerical aperture (NA) of the detector. For small NA, the maximum intensity is bounded by the Dicke-like quadratic scaling predicted in previous works~\cite{Rehler1971}, while for large apertures the dimensional scaling bound is recovered. Our results clarify a long-standing ambiguity in the relation between the scaling of the total emission rate and that inferred from directional measurements, and provide a quantitative criterion (the detection solid angle relative to the emission cone $\Delta\Omega$) that connects the two, directly guiding the design and interpretation of such experiments.

\emph{Model ---} We consider an ensemble of $N$ two-level atoms with a resonant dipole transition ($\ket{g}\!\!\leftrightarrow \!\!\ket{e}$) at frequency $\w_0=c k_0$ and linewidth $\Gamma_0$. The dynamics of the atoms interacting with the electromagnetic field in free space is described by the Markovian master equation~\cite{Lehmberg1970a}
\be\label{eq:ME}
	\partial_t \rhop = -\frac{\im}{\hbar} [\Hop,\rhop] + \sum_{i,j=1}^N \Gamma_{ij}\!\pare{\smi_i\rhop\spl_j -\inv{2} \cpare{\spl_j\smi_i,\rhop}},
\ee
where $\spl_j\equiv \ketbra{e_j}{g_j}=(\smi_j)^\dag$ is the raising operator for the $j$-th atom. $\Hop$ describes the coherent dipole-dipole interaction between atoms. As shown in Ref.~\cite{Mok2026}, adding local terms in the Hamiltonian (e.g. local fields or few-body interactions) and in the dissipator (e.g. incoherent pumping or dephasing) does not change the scaling of $R_\star$.
The symmetric $N\!\times\! N$ dissipative interaction matrix $\boldsymbol{\Gamma}$ represents the collective decay of the system. Its diagonal elements $\Gamma_{jj}\equiv\Gamma_0$ encode local decay, while the off-diagonal elements $\Gamma_{ij} \equiv 6\pi k_0^{-1}\Gamma_0 \hat{\dpp}^* \Im[\bold{G}(\rr_i-\rr_j,\w_0)]\hat{\dpp}$ (where $\bold{G}(\rr,\w_0)$ is the electromagnetic Green's tensor and $\hat{\dpp}\equiv \dpp/|\dpp|$ is the direction of atomic polarization) encapsulate the dissipative coupling between atoms $i$ and $j$~\cite{Lehmberg1970a, AsenjoGarcia2017PRX}. We denote the largest eigenvalue of $\boldsymbol{\Gamma}$ as $\Gamma_\text{max}$.

The intensity of the radiated light within a solid angle $\text{d}\uu \equiv \text{d}\phi\text{d}\theta \sin\theta$ along a direction $\uu=(\cos\phi\sin\theta,\sin\phi \sin\theta,\cos\theta)^T$, with azimuthal $\phi\in[0,2\pi]$ and polar $\theta\in [0,\pi]$ angles, is given by $I(\uu,t)\text{d}\uu$ where~\citep[pp.~178]{AllenEberlyBook}
\be\label{eq:I_n}
    I(\uu,t) = \hbar \w_0 \Gamma_0\mathcal{D}(\uu)\sum_{i,j=1}^N e^{\im k_0 \uu \cdot(\rr_i-\rr_j)} \avg{\spl_i\smi_j},
\ee
and the expectation value is taken on the state of the system $\rhop(t)$ evolved according to \eqnref{eq:ME}. Here, $\mathcal{D}(\uu)\equiv (3/8\pi)\sum_{\boldsymbol{\veps}}  |\hat{\dpp}\cdot \boldsymbol{\veps}|^2$ is the dipole emission pattern, normalized such that $\int \!\!\text{d}\uu\, \mathcal{D}(\uu)=1$. The total radiated power, obtained by integrating Eq.~(\ref{eq:I_n}) over all directions, reads
\be\label{eq:I_integrated}
	I(t) \equiv\!\! \int\!\!\text{d}\uu\,I(\uu,t) \!=\! \hbar \w_0 \!\mean{\!\sum_{i,j=1}^N \Gamma_{ji}\spl_i\smi_j\!}\! \equiv \hbar \w_0 R(t),
\ee 
where $\Gamma_{ji} = \Gamma_0 \int\!\!\text{d}\uu\,\mathcal{D}(\uu) \exp[\im k_0\uu\cdot(\rr_i-\rr_j)]$~\cite{Lehmberg1970a}. In the last step, we used \eqnref{eq:ME} to show that 
$\text{d} \avg{\hat{n}_\text{exc}}/\text{d}t = -\avg{\sum_{i,j=1}^N \Gamma_{ji}\spl_i\smi_j} $, where $\hat{n}_\text{exc}\equiv\sum_j \spl_j\smi_j$, and defined the decay rate $R(t) \equiv -\text{d} \avg{\hat{n}_\text{exc}}/\text{d}t$. 

\emph{Bounds on photon emission ---} The maximum decay rate is defined as 
\be\label{eq:I_star}
    R_\star \equiv \max_{\ket{\psi}}\bra{\psi} \frac{\Hop_\Gamma}{\hbar} \ket{\psi}, \,\text{with }\, \Hop_\Gamma \equiv \hbar \sum_{i,j=1}^N \Gamma_{ji}\spl_i\smi_j,
\ee
where the maximum is taken over all states in the Hilbert space. Obtaining $R_\star$ exactly is generally expected to be a computationally hard problem, since it is equivalent to finding the ground state energy of the spin Hamiltonian $-\hat{H}_{\Gamma}$~\cite{Kempe2006complexity} (note that $\hat{H}_\Gamma$ is not the physical Hamiltonian). Reference~\cite{Mok2026} obtained rigorous bounds on $R_\star$ using a product-state ansatz that restricts the optimization to a subset of the Hilbert space.

Here, we relax the product-state constraint and improve the upper bound by a constant prefactor, i.e., $\max\avg{\Hop_\Gamma}/\hbar \leq \Gamma_\text{max} \max\avg{\hat{\boldsymbol{\sigma}}^\dag\cdot\hat{\boldsymbol{\sigma}}} = N\Gamma_\text{max}$ where $\hat{\boldsymbol{\sigma}}\equiv (\smi_1,\ldots,\smi_N)^T$.
This result admits a clear physical interpretation: the maximum decay rate is limited by the largest number of excitations $N$ scaled by the  maximum rate at which the system can emit a single excitation. We show below that this bound is asymptotically tight and can be saturated in specific physical regimes. 

Thus $R_\star$ must satisfy the following inequalities 
\be\label{eq:Bounds}
	\max\cpare{N\Gamma_0,\frac{\Gamma_\text{max}}{4}\norm{\boldsymbol{\psi}_\text{max}}_1^2}\leq R_\star  \leq  N\Gamma_\text{max}.
\ee
The derivation of the lower bound is summarized for completeness at the end of this Letter (End Matter). It depends on the L1-norm $\norm{\boldsymbol{\psi}_\text{max}}_1^2 = (\sum_j |\psi_\text{max}^j|)^2$, where $\boldsymbol{\psi}_\text{max}$ is the normalized (in the sense of the L2-norm) principal eigenvector of $\boldsymbol{\Gamma}$~\footnote{The normalization is intended in the usual L2-norm, $\norm{\boldsymbol{\psi}_\text{max}}_2^2\equiv \sum_j |\psi_\text{max}^j|^2=1$}. The L1-norm quantifies the delocalization of the collective jump operator associated with the fastest decay channel $\cop_1 \equiv \sum_j \psi_\text{max}^j \smi_j$.

A scaling law for $R_\star$ can be derived when the brightest jump operator is delocalized over the whole system, $\norm{\boldsymbol{\psi}_\text{max}}_1^2\sim N$, so that the upper and lower bounds in \eqnref{eq:Bounds} are asymptotically tight. As full delocalization does not follow from the eigenvalue scaling alone [End Matter], proving a scaling law therefore requires the additional condition $\norm{\boldsymbol{\psi}_\text{max}}_1^2 \sim N$.
For a $D$-dimensional atomic array ($D=1,2,3$) in free space the lattice periodicity ensures this property. Because for ordered arrays $\Gamma_\text{max} \sim N^{\frac{1}{2}-\frac{1}{2D}}\Gamma_0$, Eq.~\eqref{eq:Bounds} leads to a \emph{dimensional scaling}~\cite{Mok2026} that depends only on the lattice dimensionality. Crucially, this argument seems to imply the necessity of crystalline order, which makes the scaling appear as a special case requiring fine tuning.

\begin{figure}
	\includegraphics[width=\columnwidth]{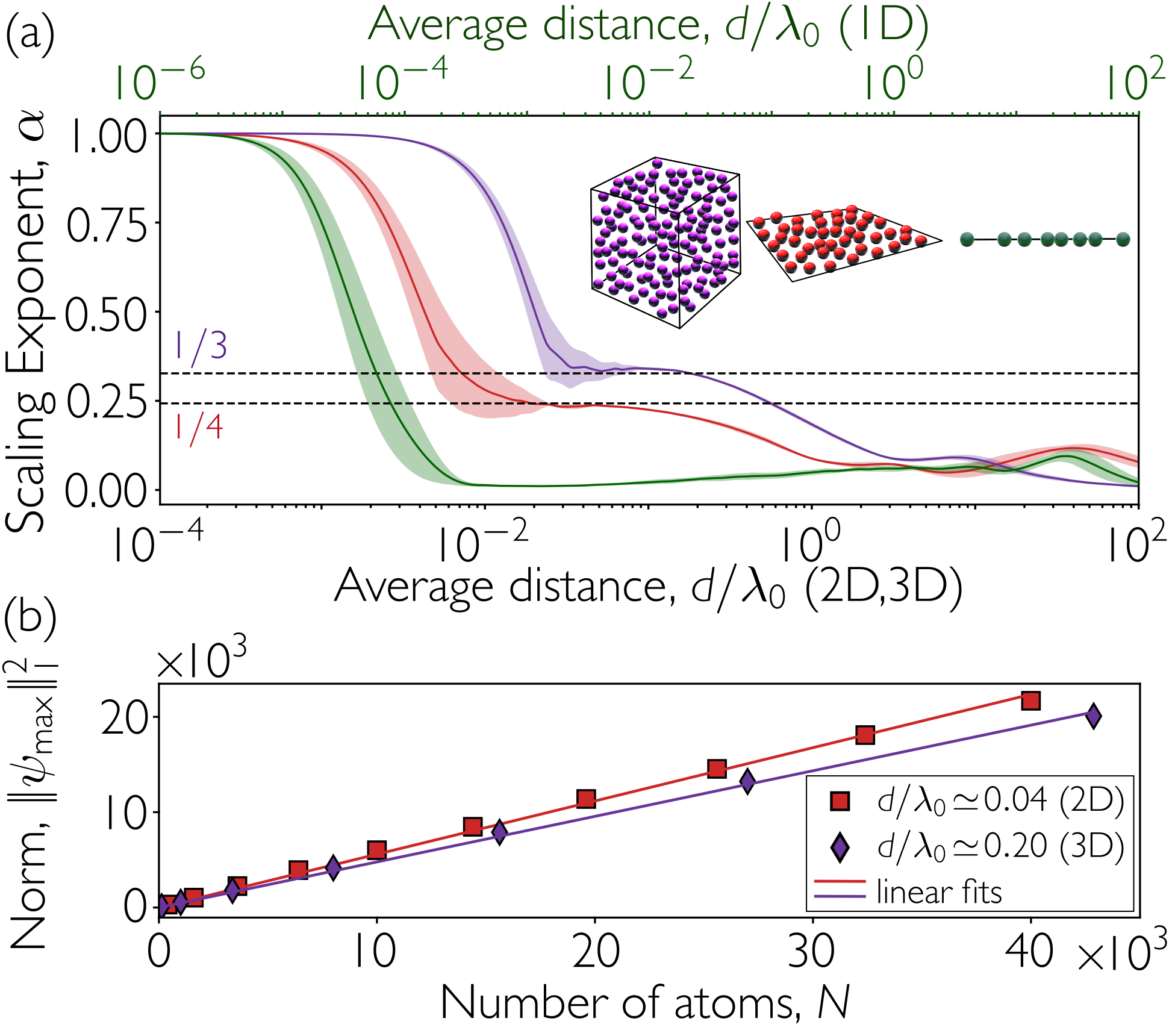}
	\caption{(a) Scaling exponent obtained from a fit of $\Gamma_\text{max}$ to $\beta N^\alpha \Gamma_0$, as a function of the average interatomic separation $d/\lambda_0$ for 1D, 2D, and 3D atomic clouds. The colored region represents the $1\sigma$ confidence interval. The fits are done over a region $N_\text{1D}\in[N_\text{min},N_\text{max}]$ where $N_\text{min}=\{3000,20\}$ and $N_\text{max}=\{3\times 10^4,200\}$ in steps of $\{3\times10^3,20\}$ for $D=\{1,2\}$ respectively and for $D=3$ we used $N_\text{1D}=\{5, 8, 10, 13, 15, 18, 20, 23, 25, 28, 30, 35\}$.  (b) Plot of $|\!|\boldsymbol{\psi}_\text{max}|\!|_1^2$ as a function of atom number, for 2D (red squares) and 3D (purple diamonds) clouds with $d/\lambda_0 \simeq 0.04$ and $d/\lambda_0 \simeq 0.2$ respectively. The solid lines represent linear fits. In both panels, the results are obtained by averaging over 500 (2D) and 100 (1D and 3D) realizations.}\label{fig:Fig2}
\end{figure}

\emph{Scaling law for disordered ensembles ---} We now show that the \textit{dimensional scaling} for ensembles at constant density
\be\label{eq:Scaling_Law}
    R_\star \sim N\Gamma_\text{max} \sim N^{\frac{3}{2}-\frac{1}{2D}}\Gamma_0 ,
\ee
holds more generally for dense but extended clouds of atoms with linear dimensions $L$ larger than the characteristic dipole transition wavelength $\lambda_0$. 
For a gas of atoms, $\boldsymbol{\Gamma}$ is a Euclidean random matrix~\cite{Mezard1999} ---with entries obtained by evaluating $\Gamma_{ij}$ at randomly sampled atomic positions--- and its spectral properties have been studied extensively in previous work~\cite{Akkermans2008, Skipetrov2011, Gero2013, Goetschy2013, Viggiano2023, Viggiano2025}. 
However, the scalings of $\Gamma_\text{max}$ and $\norm{\boldsymbol{\psi}_\text{max}}_1^2$ with $N$ are, to the best of our knowledge, only known for dense  3D clouds~\cite{Svidzinsky2008}, a gap that we fill here.

We solve the eigenvalue problem of $\boldsymbol{\Gamma}$ in the limit of large atomic densities, where superradiance is expected to arise. In the large density limit, the spectrum of $\boldsymbol{\Gamma}$ is asymptotically close to the spectrum of the integral operator~\cite{Koltchinskii2000, Williams2000}
\be\label{eq:Fredholm_Eq}
	\int_{\mathbb{R}^D}\!\!\text{d}\rr'\, \rho_D(\rr') \Gamma(\rr,\rr') \psi_n(\rr') = \Gamma_n\psi_n(\rr),
\ee
where $\rho_D(\rr)$ is the density profile of the $D$-dimensional gas of atoms, $\psi_n(\rr)$ the eigenfunction of the integral operator and $\Gamma_n$ its associated eigenvalue.
We consider the scalar model for the dissipative matrix, 
\be\label{eq:Gamma_kernel_scalar}
    \Gamma(\rr,\rr') = \Gamma_0\frac{\sin(k_0|\rr-\rr'|)}{k_0|\rr-\rr'|},
\ee
which is obtained from the general form of $\Gamma_{ij}$ by setting $\mathcal{D}(\uu)=1/4\pi$ and describes a random gas of unpolarized atoms.
We also assume atoms to be distributed in a $D$-dimensional ball with uniform density such that $\rho_\text{1D} \equiv (N/L)\delta(z)\delta(y)$, $\rho_\text{2D} \equiv (N/\pi L^2)\delta(z)$, and $\rho_\text{3D} \equiv 3N/4\pi L^3$.
These simplifying assumptions allow us to derive analytical results for $\Gamma_\text{max}$ and $\boldsymbol{\psi}_\text{max}$. 
The same qualitative behavior is obtained including the full tensorial nature of the Green's function~\cite{Bellando2021} and different atomic distributions, as shown numerically in the Supplemental Material (SM)~\footnote{See the Supplemental Material for: (i) additional information about the fitting procedure and the numerical simulations; (ii) the analytical derivations of the scaling of $\Gamma_\text{max}$ in 1D, 2D, and 3D clouds; (iii) the proof that \eqnref{eq:Ansatz_Eigenstates} are eigenfunctions of the integral equations; (iv) derivation of the angular distribution of photon emission from ordered and disordered ensembles; (v) scaling of $R_\star$ using semidefinite programming relaxation (SDP); (vi) discussion of other electromagnetic environments. The Supplemental Material includes Refs.~\expandafter\cite\expandafter{\SMreflist}\label{footnote: SM}}\expandafter\nocite\expandafter{\SMreflist}%
\newcounter{firstfootnote}
\setcounter{firstfootnote}{\value{footnote}}.

We proceed in two steps: (i) we prove that $\Gamma_\text{max}$ has the same scaling as for ordered arrays and (ii) we show that the principal eigenfunction is delocalized, i.e.,  $\norm{\boldsymbol{\psi}_\text{max}}_1^2\sim N$. The largest eigenvalue of \eqnref{eq:Fredholm_Eq} can be obtained using Gelfand's formula for the spectral radius of positive operators, namely $\Gamma_\text{max} = \lim_{m\rightarrow \infty} \sqrt[m]{\Tr[\Gamma^m]}$, where $\Tr[\Gamma^2]=\int\!\!\text{d}\bold{x}\text{d}\bold{y} \rho_D(\bold{x})\rho_D(\bold{y})\, \Gamma(\bold{x},\bold{y})\Gamma(\bold{y},\bold{x})$, and yields the scaling $\Gamma_\text{max}\sim N^{\frac{1}{2}-\frac{1}{2D}} \Gamma_0$ (see SM~\footnotemark[\value{firstfootnote}]).
For the eigenvectors of \eqnref{eq:Fredholm_Eq}, we consider the ansatz
\be\label{eq:Ansatz_Eigenstates}
	\psi_\nn(\rr) = \mathcal{N}_n 
	\left\{
	\begin{array}{ll}
	j_n(k_0r)& \quad \text{for $D=1$},\\
	J_n(k_0r)e^{\im n\phi}& \quad \text{for $D=2$},\\
	j_n(k_0 r) Y_{nm}(\theta,\phi) &\quad \text{for $D=3$},
	\end{array}
	\right .
\ee
where the eigenvectors are labeled by the single index $n$ for 1D and 2D, and by two indices ${n,m}$ for 3D; $\mathcal{N}_n$ is a normalization constant, $J_n(k_0r)$ and $j_n(k_0r)$ are, respectively, Bessel and spherical Bessel functions of the first kind and order $n$, and $Y_{nm}(\theta,\phi)$ are spherical harmonics of order $n$ and $m$. We note that $n$ goes up to $ \lfloor k_0L \rfloor$ in 2D ($\lfloor \sqrt{k_0L} \rfloor$ in 1D) where $\lfloor\cdot \rfloor$ is the floor function~\footnotemark[\value{firstfootnote}]. Accordingly, \eqnref{eq:Ansatz_Eigenstates} only captures a subset of eigenvectors of \eqnref{eq:Fredholm_Eq} for 1D and 2D ensembles. From the properties of the Bessel functions we interpret $n$ as quantifying the distribution of atomic excitations, with smaller values indicating bulk excitations, and excitations closer to the boundary for $n\sim\lfloor k_0L\rfloor$.
Substituting \eqnref{eq:Ansatz_Eigenstates} into \eqnref{eq:Fredholm_Eq} and decomposing the kernel into a sum of spherical waves, we show in the SM~\footnotemark[\value{firstfootnote}] that \eqnref{eq:Ansatz_Eigenstates} is a good approximation for some of the eigenvectors of \eqnref{eq:Fredholm_Eq} ---the ones associated with largest eigenvalues--- in the limit of large system size $L/\lambda_0\gg1$. For 3D clouds, \eqnref{eq:Ansatz_Eigenstates} is exact and forms a complete set~\cite{Svidzinsky2008}. For $n^2\ll k_0 L$, the eigenvalues associated with \eqnref{eq:Ansatz_Eigenstates} have the same scaling as $\Gamma_\text{max}$, and for 2D and 3D clouds $\norm{\psi_n(\rr)}_1^2 \sim N$ [End Matter]. Since for any eigenvalue-eigenvector pair $\Gamma_n \norm{\boldsymbol{\psi}_n}_1^2/4$ constitutes a valid lower bound on $R_\star$, we conclude the validity of the dimensional scaling in \eqnref{eq:Scaling_Law} for disordered atomic ensembles.
In the SM, we independently confirm the scaling law for 2D and 3D atomic clouds via a semidefinite programming relaxation (SDP) for $\Hop_\Gamma$, which yields a scaling law for $R_\star$~\cite{Mok2026}.

We now compute numerically at which atomic densities the results obtained from \eqnref{eq:Fredholm_Eq} correctly reproduce the largest eigenvalue of the Euclidean random matrix $\boldsymbol{\Gamma}$ obtained by sampling \eqnref{eq:Gamma_kernel_scalar} on the atomic distribution.
We compute the scaling of $\Gamma_\text{max}$ by diagonalizing the matrix $\boldsymbol{\Gamma}$ for different numbers of atoms $N$ at fixed density and fitting the result to the formula $\Gamma_\text{max}= \Gamma_0\beta N^\alpha$. We find that the dimensional scaling $\alpha \simeq 1/2-1/2D$ predicted by  Gelfand's formula holds for $d/\lambda_0 \approx 0.01-0.1$ in 2D and $d/\lambda_0 \approx 0.04-0.2$ in 3D [\figref{fig:Fig2}(a)]. For these values of $d/\lambda_0$, we confirm that the support of the principal eigenvector scales with system size $\norm{\psi_\text{max}}_1^2\sim N$ for both 2D and 3D clouds [\figref{fig:Fig2}(b)]. To have a sense of the atomic densities at which these results hold, we consider a 3D cloud of $^{88}$Sr atoms radiating on the closed two-level transition $^3D_3\rightarrow {}^3P_2$ at $\lambda_0\simeq 2.9\mu\text{m}$~\cite{Sansonetti2010}. For an average separation $d/\lambda_0 = 0.1$, we obtain $\rho \approx 4\times 10^{13}\text{cm}^{-3}$, which has been achieved experimentally~\cite{FrometaFernandez2025}.
The results in \figref{fig:Fig2} have been obtained assuming atoms uniformly distributed in a 1D, 2D, and 3D square box to better compare with the result for arrays. However, these findings are independent of this choice (see SM~\footnotemark[\value{firstfootnote}]).

The exact position and extent of the dimensional-scaling region depend on both the maximum system size $N$ used in the fit, with larger systems leading to larger regions, as well as on the total span of values for $N$. In \figref{fig:Fig2}, we consider a large span of values $N$ from moderate to large system sizes. In the SM, we analyze the experimentally relevant case in which the fit is done for a handful of values around a specific size $N$ and show qualitatively similar results. 

The scaling of $\Gamma_\text{max}$ as a function of $d/\lambda_0$ in \figref{fig:Fig2}(a) is qualitatively the same as for arrays, confirming a universal behavior for the scaling exponent $\alpha$.  
As for arrays, the expected Dicke scaling ($\alpha=1$) for $d/\lambda_0\rightarrow 0$ transitions to the dimensional scaling $\alpha = 1/2-1/2D$ at $L = dN^{1/D}\sim \lambda_0$. As a function of $d/\lambda_0$, the transition region thus depends sensitively on $N$, as reflected in the larger uncertainty in \figref{fig:Fig2}(a).
The convergence to the non-interacting limit ($\alpha=0$) for $d/\lambda_0\gg 1$ is less obvious than for arrays. At large average interatomic distance, two atoms can still be within a distance smaller than $\lambda_0$, at which point their decay can be collectively enhanced. In this regime, $\Gamma_\text{max}$ is thus determined by superradiant pairs of atoms and its value is distributed in $[\Gamma_0,2\Gamma_0]$~\cite{Andreoli2021}. This interval is independent of $N$, and one would thus expect to obtain $\alpha=0$. However, for fixed density, values close to $2\Gamma_0$ become more probable for larger $N$, leading to a fit $\alpha\neq 0$, which means the model $\beta N^\alpha$ is a poor fit for the results (see SM~\footnotemark[\value{firstfootnote}]). This effect is more pronounced in lower dimensions, as the probability for two atoms to lie within a distance $d_c$ scales as $(d_c/L)^D$ and is thus larger for smaller $D$.

\begin{figure}
	\includegraphics[width=\columnwidth]{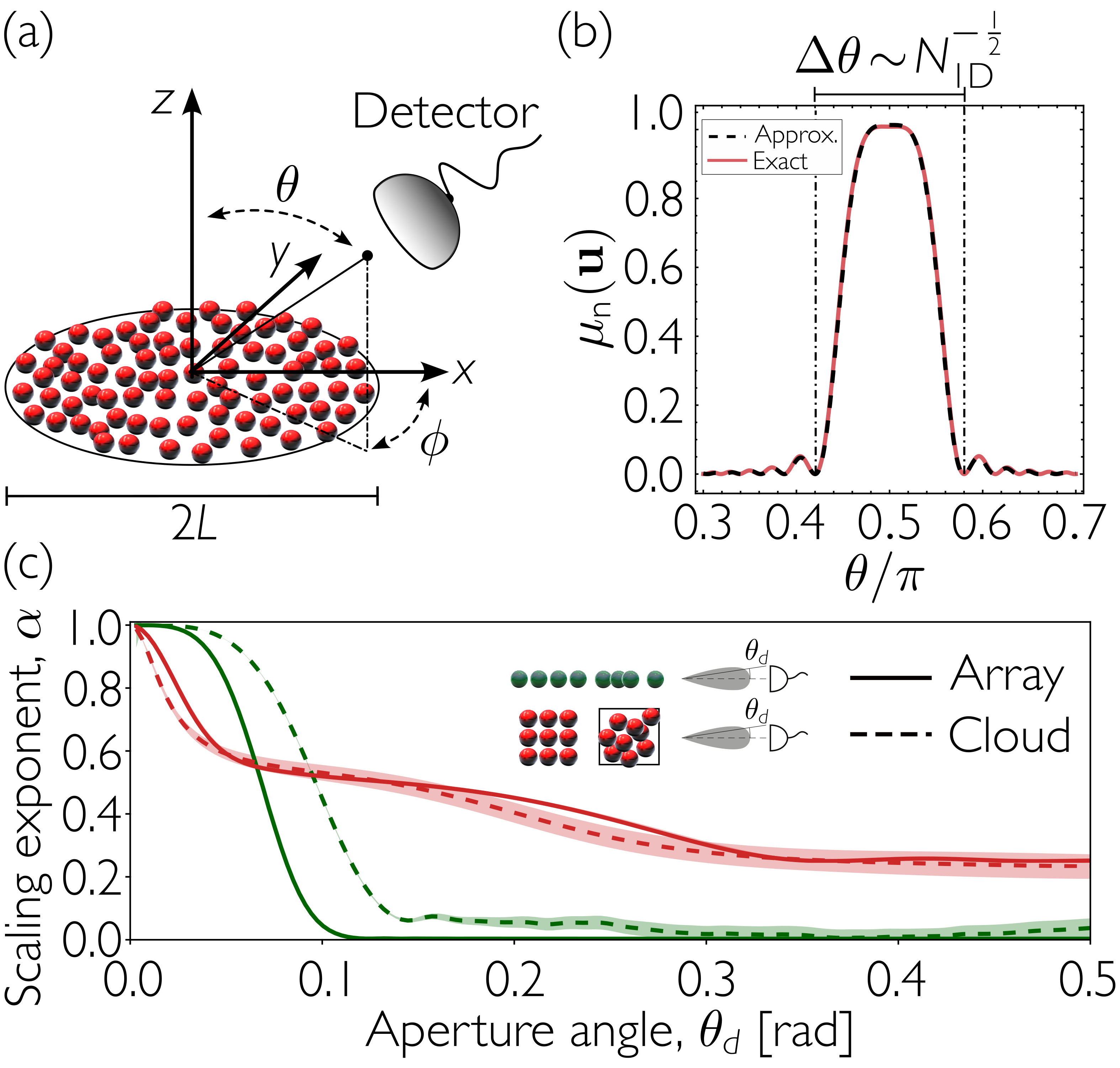}
	\caption{(a) Schematic of directional detection of photons emitted by atoms uniformly distributed on a disk of radius $L$. (b) Emission profile of the dominant eigenstate $\psi_0(\rr)$, defined in \eqnref{eq:Ansatz_Eigenstates}, for $k_0L=100$. The emission is symmetric about the disk axis and collimated about the plane of the disk within an angle $\Delta\theta\!\sim \!(k_0L)^{-1/2}$. Numerical evaluation of Eq.~\eqref{eq:I_directional_k0} (solid red) compared to the approximation in \eqnref{eq:mu_approx} (black dashed). (c) Scaling of $\Gamma_\text{max}^\text{NA}$ with $N$ (obtained from a best fit to $\beta N^\alpha \Gamma_0$) as a function of aperture angle $\theta_d$ for both 1D (green) and 2D (red) ensembles. Solid (dashed) lines correspond to arrays (clouds). For one- (two-) dimensional arrays atoms are polarized perpendicular (parallel) to the array direction, and $d=0.5\lambda_0$. The fits are done for ensembles with even number of atoms $N_{\text{1D}}\in[900,1000]$ for 1D, and $N_{\text{1D}}\in[40,50]$ for 2D. For clouds the result is obtained averaging over 100 realizations.}\label{fig:Fig3}
\end{figure}

\textit{Connection with optical depth ---} The fact that Eq.~\eqref{eq:Scaling_Law} holds for both ordered arrays and disordered clouds suggests that its origin transcends the microscopic details of the distribution. We now show that the relevant geometric parameter is the optical depth: $\Gamma_\text{max}/\Gamma_0 \sim \text{OD}$. Following Refs.~\cite{Rehler1971,Chang2018,TanjiSuzuki2011}, we use the geometric definition $\text{OD}\sim N\Delta\Omega$, where $\Delta\Omega$ is the solid angle into which the brightest collective mode radiates. We note that this geometric OD differs from the probe-attenuation OD measured at high density, which is reduced by coherent dipole-dipole interactions~\cite{Andreoli2021}; the bound on $R_\star$ is governed by the geometric quantity, set by the spectrum of $\boldsymbol{\Gamma}$. To make apparent the connection between $\Gamma_\text{max}$ and the OD, we derive the scaling of $\Delta\Omega$ with the size of the atomic ensemble when the density is kept fixed. 

For the single-excitation eigenstates  in \eqnref{eq:Ansatz_Eigenstates}, the directional emitted intensity \eqnref{eq:I_n} reads
\be\label{eq:I_directional_k0}
\begin{split}
	I_n(\uu,t) =& \hbar \w_0 \Gamma_0 N \mathcal{D}(\uu) \mu_n(\uu),
\end{split}
\ee
where the effects of interference between the dipoles on the emission profile are encapsulated by the diffraction function 
\be\label{eq:mu_factor_def}
	\mu_n(\uu) \equiv \bigg\vert \inv{\sqrt{N}}\!\int\!\! \text{d}\rr\, \rho_\text{D}(\rr) e^{-\im k_0 \uu \cdot\rr} \psi_n(\rr)\bigg\vert^2.
\ee
Here, $\mu_n(\uu)$ includes both independent and collective contributions, whereas previous work defined it as the emission pattern arising solely from collective decay~\cite{Rehler1971}.

Let us compute \eqnref{eq:mu_factor_def} for the case of $N$ atoms arranged in a disk of radius $L\equiv N_\text{1D} d$ with uniform density $\rho_\text{2D}(\rr)=N/\pi L^2$ [\figref{fig:Fig3}(a)]. Using properties of Bessel functions, the integral in \eqnref{eq:mu_factor_def} can be solved exactly. Due to the cylindrical symmetry of the ensemble, the resulting angular profile of the emission is independent of the azimuthal angle $\phi$ and is mostly confined to the plane of the array [\figref{fig:Fig3}(b)]. In the large ensemble limit, for $n\ll k_0L$, and expanding $\theta$ around $\pi/2$, we obtain 
\be\label{eq:mu_approx}
    \mu_n(\uu) \simeq \frac{2}{\pi k_0 L}\pare{\frac{\sin(k_0L \tilde{\theta}^2/2)}{k_0L\tilde{\theta}^2/2}}^2,
\ee
where $\tilde{\theta} \equiv \theta - \pi/2$.
We find excellent agreement between \eqnref{eq:mu_approx} and the exact result for the emission profile [\figref{fig:Fig3}(b)]. 
From \eqnref{eq:mu_approx} we estimate that emission in the plane perpendicular to the disk is collimated within an angle $\Delta\theta = 2\sqrt{2\pi/k_0L}$, defined as the distance between the first two zeros of $\mu_n(\uu)$ in \figref{fig:Fig3}(b).
The probability of emitting a photon along a direction $\phi$ in the $xy$-plane is, instead, uniform.
However, when a photon is detected along a particular direction $\phi$, it can be concluded that it was emitted within a diffraction angle $\Delta\phi\sim 1/k_0L$, determined by the length of the perimeter of the disk, akin to diffraction from a line of scatterers of length $L$. Accordingly, photons are emitted within a solid angle $\Delta\W \sim \Delta\phi\Delta\theta \sim (k_0L)^{-3/2}$, which yields OD $\sim \sqrt{k_0L}$.
We remark that this result, here proven for 2D ensembles, holds for both arrays and clouds of any dimensionality (SM~\footnotemark[\value{firstfootnote}]).
This establishes the connection between $R_\star$ and OD and concludes the proof of \eqnref{eq:Scaling_Law_OD}.

\emph{Directional scaling ---} The angular dependence of the emission profile suggests that the scaling of the emitted intensity is a function of the numerical aperture of the detector. While for large NA the maximum intensity is given by \eqnref{eq:Scaling_Law}, constructive interference can lead to a different, enhanced scaling for the detected intensity if only a small fraction of the light is detected. 
Analogously to Eq.~\eqref{eq:I_star}, we define the maximum decay rate measured by a detector oriented in a direction $\uu$ as
\be\label{eq:Directional_Scaling_Law}
    R_\star^{\text{NA}} \equiv \max_{\ket{\psi}}\bra{\psi} \frac{\Hop_\Gamma^{\text{NA}}}{\hbar} \ket{\psi}, \text{with }\; \Hop_\Gamma^{\text{NA}} \equiv \hbar\sum_{i,j=1}^N \Gamma_{ji}^{\text{NA}}\spl_i\smi_j.
\ee
Here, $\Gamma_{ji}^\text{NA} \equiv \Gamma_0 \int_{\text{NA}}\!\!\text{d}\uu\,\mathcal{D}(\uu) e^{\im k_0\uu\cdot(\rr_i-\rr_j)}$ is the ``directional dissipative matrix'', of largest eigenvalue $\Gamma_\text{max}^\text{NA}$.  The same bounding argument as for Eq.~\eqref{eq:I_star} yields $R_\star^\text{NA} \leq N\Gamma_\text{max}^\text{NA}.$

Since $[\Hop_\Gamma^{\text{NA}},\Hop_\Gamma]\neq 0$, directional and total emission are governed by different operators with different optimal states. In the limit of large aperture angle, \eqnref{eq:Directional_Scaling_Law} reduces to \eqnref{eq:I_star}, with the scaling of \eqnref{eq:Scaling_Law}. For a small solid angle $\text{d}\uu$ around the direction $\uu$,  $\Gamma_{ji}^\text{NA} \simeq \Gamma_0 \text{d}\uu D(\uu) e^{\im k_0\uu\cdot(\rr_i-\rr_j)}$. In this limit, the principal eigenvector reads $\boldsymbol{\psi}_{\text{max}} = (e^{\im k_0 \uu\cdot \rr_1},\ldots,e^{\im k_0 \uu\cdot \rr_N})^T/\sqrt{N}$ and $\Gamma_\text{max}^\text{NA} = \Gamma_0 \text{d}\uu D(\uu) N$. This result holds for generic ensembles of any dimensionality. We confirm it numerically by computing the scaling of $\Gamma_\text{max}^\text{NA}$ as a function of the detector aperture angle for 1D and 2D arrays and clouds in \figref{fig:Fig3}(c). 

These results show that for sufficiently small NAs, the maximal intensity always scales as $N^2$ because of constructive interference in the emission. This scaling holds for any direction $\uu$, except those forbidden by atomic polarization (i.e., when $D(\uu)=0$). In practice, superradiant emission proceeds preferentially through the brightest modes of $\Hop_\Gamma$~\cite{Clemens2003, Masson2020, Masson2022}, so detection along directions that do not overlap with these modes, while possible, is statistically unlikely~\cite{Ernst1968,Clemens2003}.

In conclusion, we showed that the maximum emission rate from an atomic ensemble obeys the universal scaling law $R_\star \sim \Gamma_0 N \times \OD$, unifying ordered arrays and disordered clouds under a single framework governed by the optical depth. In the SM, we extend these conclusions to encompass atoms in cavities and along waveguides.
This bound thus extends beyond free space to any system whose light-matter coupling is governed by the optical depth, with immediate consequences for driven-dissipative phase transitions~\cite{Narducci1978,Carmichael1980,Hepp1973} and superradiant lasing~\cite{Meiser2009,Bohnet2012}.
Our results also provide a rigorous benchmark for approximate methods~\cite{Ostermann2024, Agarwal2024, Goncalves2024, Ruostekoski2024, Mink2023}. The connection between OD and emission angle further reveals that directional detection can yield a different scaling: for sufficiently small detector angles, $R_\star^\text{NA} \sim N^2$, a Dicke-like behavior that calls for care in interpreting directional measurements~\cite{Ferioli2023, Liedl2024}.

Finally, whether a state decaying at $R_\star$ can be dynamically prepared from the fully inverted state $\ket{e}^{\otimes N}$ remains open. Inhomogeneous broadening from dipolar interactions is particularly severe in gases and can affect the scaling of the effective optical depth~\cite{He2020, Andreoli2021, Grava2022, Ji2023}, potentially suppressing the cooperative phenomena underlying superradiant bursts~\cite{Gross1982}. Lattice trapping mitigates broadening in arrays~\cite{Masson2022, Sierra2022, Robicheaux2021}. Nevertheless, the product state $\bigotimes_j(\ket{e}+e^{i\kk_0\cdot \rr_j}\ket{g})/\sqrt{2}$ decays at a rate $\propto R_\star$ for both arrays~\cite{Mok2026} and clouds [End Matter], and can be prepared with a sufficiently strong drive, rendering dipolar interactions negligible compared to the drive during the preparation. The scaling $R_\star$ can thus be determined by preparing this state at fixed density and increasing $N$, monitoring either the excited population with a quantum gas microscope~\cite{Douglas2026} or the emitted light with an NA large enough to capture the emission cone (End Matter).

\emph{Acknowledgements ---} C.C.R. acknowledges support from the European Union’s Horizon Europe program under the Marie Sk\l{}odowska Curie Action LIME (Grant No. 101105916). We acknowledge support by the National Science Foundation through the CAREER Award (No. 2047380), the Air Force Office of Scientific Research through their Young Investigator Program (grant No. 21RT0751), the U.S. Department of Energy, Office of Science, National Quantum Information Science Research Centers, Co-design Center for Quantum Advantage (C2QA) under Contract No.~DE-SC0012704, as well as by  Chu Family Foundation and the David and Lucile Packard Foundation. A.A.-G.~also acknowledges support from Programmable Quantum Materials, an Energy Frontier Research Center funded by the U.S. Department of Energy, Office of Science, Basic Energy Sciences (BES), under Award No.~DE-SC0019443. S. B. J. acknowledges support by the Deutsche Forschungsgemeinschaft (DFG, German Research Foundation) under Projects No. 277625399-TRR 185 OSCAR (`` Open System Control of Atomic and Photonic Matter'', B4), No. 277146847-CRC 1238 (``Control and Dynamics of Quantum Materials'', C05), and under Germany’s Excellence Strategy-Cluster of Excellence Matter and Light for Quantum Computing (ML4Q) EXC 2004/1–390534769. A.P. acknowledges support from the Harvard Purcell fellowship.
W.K.M acknowledges funding provided by the Institute for Quantum Information and Matter, an NSF Physics Frontiers Center (NSF Grant PHY-2317110).

\emph{Data availability ---} numerical data and codes for the main figures are available at \cite{zenodo_dataset}.

\let\oldaddcontentsline\addcontentsline
\renewcommand{\addcontentsline}[3]{}
\bibliography{Bibliography}
\let\addcontentsline\oldaddcontentsline

\begin{center}
\vspace{1cm}
\textbf{\large End Matter}\\[.2cm]
\end{center}

{\bf Derivation of the lower bound in \eqnref{eq:Bounds}} --- For completeness, we reproduce the derivation of the lower bound in \eqnref{eq:Bounds} originally presented in Ref.~\cite{Mok2026}. First, we notice that the variational method yields the simple lower bound $N\Gamma_0$ for the state $\ket{e}^{\otimes N}$. In the presence of dissipative interactions ($\Gamma_{ij}\neq 0$) this bound can be tightened. We introduce the collective jump operators $\cop_n = \boldsymbol{\psi}_n\cdot \boldsymbol{\smi}$, where $\boldsymbol{\psi}_n$ is the eigenvector of $\boldsymbol{\Gamma}$ with eigenvalue $\Gamma_n$. In terms of these jump operators, $\Hop_\Gamma = \sum_n \Gamma_n \cdop_n\cop_n$.
Noting that $\boldsymbol{\Gamma}\succeq 0$, the maximum emission rate thus reads 
\be\label{eq:generic_lower_bound}
    R_\star = \max_{\ket{\psi}} \bra{\psi}\frac{\Hop_\Gamma}{\hbar}\ket{\psi} 
    \geq \Gamma_n \max_{\ket{\psi}} \bra{\psi} \cdop_n\cop_n \ket{\psi} 
    = \Gamma_n \norm{\cdop_n\cop_n},
\ee
which is valid for all choices of $n$. We now need to compute the norm on the right hand side of \eqnref{eq:generic_lower_bound}. We have
\be
\begin{split}
    \norm{\cdop_n\cop_n}\! =& \Big\Vert\sum_{ij}\psi_n^j (\psi_n^i)^* \spl_i \smi_j\Big\Vert
    \!=\! \Big\Vert\sum_{ij}|\psi_n^j| |\psi_n^i| \tilde{\sigma}^+_i \tilde{\sigma}^-_j\Big\Vert\\
    =& \max_{\ket{\phi}} \sum_{ij}|\psi_n^j| |\psi_n^i| \avg{\tilde{\sigma}^+_i \tilde{\sigma}^-_j} \geq \inv{4}\bigg(\sum_j |\psi_n^j|\bigg)^2,
\end{split}
\ee
where we have absorbed the phases $\phi_i$ of $\psi_i$ into the lowering operators $\tilde{\sigma}_i^- = e^{i\phi_i} \hat{\sigma}_i^-$. 
Since \eqnref{eq:generic_lower_bound} is valid for any choice of eigenvalue and eigenvector, a tighter bound is obtained maximizing over $n$, which leads to \eqnref{eq:Bounds}.

{\bf Relation between scalings of $\Gamma_\text{max}$ and $\norm{\psi_n(\rr)}_1^2$} --- Let us now demonstrate that full delocalization condition $\norm{\psi_n(\rr)}_1^2\sim N$ is an independent condition that does not follow directly from the scaling of $\Gamma_\text{max}$.
For $\Gamma_\text{max}=\beta N^\alpha \Gamma_0$ ($0\leq \alpha\leq 1$), the bound $|\Gamma_{ij}|\leq\Gamma_0$ implies $\Gamma_\text{max}\leq \Gamma_0 \norm{\psi_\text{max}}_1^2$, hence $ \beta N^\alpha \leq \norm{\boldsymbol{\psi}_\text{max}}_1^2$. The Cauchy-Schwarz inequality gives the upper bound $\norm{\boldsymbol{\psi}_\text{max}}_1^2\leq N$. Substituting the lower bound into \eqnref{eq:Bounds} yields $R_\star \geq \beta^2 N^{2\alpha} \Gamma_0/4$, which for $\alpha < 1$ does not match the upper bound $N\Gamma_\text{max} \sim N^{1+\alpha}\Gamma_0$. Proving the dimensional scaling therefore requires  the stronger condition $\norm{\boldsymbol{\psi}_\text{max}}_1^2 \sim N$. 

{\bf Calculation of $\norm{\psi_n(\rr)}_1$ for dense clouds} --- In the following, we prove the asymptotic tightness of the lower bound in \eqnref{eq:Bounds} by computing  $\norm{\psi_n(\rr)}_1 = \int\!\! \text{d}\rr\, \rho_\text{D}|\psi_n(\rr)|$ where $\psi_n(\rr)$ is given by the ansatz state in \eqnref{eq:Ansatz_Eigenstates} and we assume uniform atomic density. In the following, we explicitly consider the case of a two dimensional ensemble. A similar procedure leads to the same conclusion for 3D clouds. 

In the limit of extended clouds $k_0L\gg n^2$ the normalization factor reads $\mathcal{N}_n\simeq \sqrt{k_0/2L\rho_\text{2D}}$ to leading order in $k_0L$. We find
\be\label{eq:Norm_1_psi_step1}
    \norm{\psi_n(\rr)}_1 = \sqrt{\frac{2\pi^2 \rho_\text{2D}}{k_0^3 L}}\spare{\int_0^{\eta}\!\!\!\!\text{d}x\, x|J_n(x)| + \!\int_{\eta}^{k_0L}\!\!\!\!\text{d}x\, x|J_n(x)|},
\ee
where we introduced $x=k_0r$ and defined the large constant $\eta\gg n$ that does not depend on $k_0L$. In the limit $k_0 L \gg1$, \eqnref{eq:Norm_1_psi_step1} is dominated by the second integral in the square brackets. 
For $\eta\gg n$ we can approximate $J_n(x)\simeq \sqrt{2/\pi x}\cos[x-\pi(2n+1)/4]$ and lower bound the integral as
\be\label{eq:Integral_lower_bound}
    \int_{\eta}^{k_0L}\!\!\!\!\text{d}x\, x|J_n(x)| \geq \sqrt{\frac{2}{\pi}}\int_{\eta}^{k_0L}\!\!\!\!\text{d}x\, \sqrt{x} \cos^2\pare{\varphi_n - x}
\ee
where $\varphi_n\equiv \pi(2n+1)/4$.
Using that $\cos^2\pare{\varphi_n - x} = [1+ \cos \pare{2\varphi_n - 2x}]/2$, \eqnref{eq:Integral_lower_bound} is evaluated as the sum of two integrals; one proportional to $\sqrt{x}/2$ and the other containing an oscillating contribution. The former dominates the integral scaling as $ (k_0L)^{3/2}$. Substituting this result into \eqnref{eq:Norm_1_psi_step1} the L1-norm is asymptotically lower bounded as
\be
    \norm{\psi_n(\rr)}^2_1 \geq  \frac{4\pi \rho_\text{2D}}{9k_0^2} (k_0L)^2 \sim N.
\ee
Finally, using the inequality $\norm{\cdot}_1 \leq \sqrt{N}\norm{\cdot}_2$ together with the fact that $\psi_n(\rr)$ is normalized we conclude that $\norm{\psi_n(\rr)}^2_1\sim N$. 

For 1D ensembles, the product state $\ket{e}^{\otimes N}$ already provides an asymptotically tight lower bound for $R_\star$. It is thus unnecessary to prove that $\norm{\psi_\text{max}}^2_1\sim N$. In fact, this is not the case as $\norm{\psi_n(\rr)}^2_1\sim (\log N)^2$. The tightest lower bound for 1D ensembles is thus given by $N\Gamma_0$ and it is obtained from the fully excited state $\ket{e}^{\otimes N}$.

{\bf A product state decaying at $R_\star$.} We now prove that the instantaneous decay rate of the product state $\ket{\Psi}=\bigotimes_j(\ket{e}+e^{ik_0 \nn\cdot \rr_j}\ket{g})/\sqrt{2}$ scales as $R_\star$. For ordered arrays this was proved in Ref.~\cite{Mok2026}. As mentioned in the main text, this state can be prepared by illuminating the ensemble with a sufficiently strong laser pulse directed along $\nn$, such that dipolar interactions are negligible during the pulse duration. For atomic clouds, we have
\be\label{eq:R_psi}
    R_\Psi  \equiv \sum_{ij} \Gamma_{ji} \bra{\Psi} \spl_i\smi_j \ket{\Psi} = \mu_{\Psi}(\nn)+\frac{N\Gamma_0}{4},
\ee
where we have introduced the function 
\be\label{eq:mu_Psi}
    \mu_{\Psi}(\nn) = \frac{\Gamma_0}{4}\!\int \!\!\text{d}\uu\, \mathcal{D}(\uu)\sum_{i,j=1}^N e^{\im k_0 (\uu-\nn)\cdot (\rr_i-\rr_j)}.
\ee
Note that the sum in \eqnref{eq:mu_Psi} includes $i=j$, and thus $\mu_{\Psi}(\nn)$ contains both the contribution of the interference between atoms ($i\neq j$) as well as half of the contribution from independent decay ($i=j$).
For disordered clouds the positions $\rr_i$ and $\rr_j$ are random variables sampled from the atomic spatial distribution. We thus average \eqnref{eq:mu_Psi} over all possible configurations and indicate the results as $\avg{\mu_\Psi(\nn)}_\text{cnf}$.

It is convenient to illustrate the calculation for the particular case of a 2D cloud of uniformly distributed atoms on a square of area $L^2$ in the $xy$ plane. We note that
\be\label{eq:exp_avg_rel}
    \avg{e^{\im k_0 (\uu-\nn)\cdot (\rr_i-\rr_j)}}_\text{cnf} =\!\!\! \prod_{\alpha=x,y}\!\!\!\pare{\frac{\sin[(u_\alpha-n_\alpha)k_0L/2]}{(u_\alpha-n_\alpha)k_0L/2}}^2.
\ee
The function in Eq.~\eqref{eq:exp_avg_rel} is sharply peaked around the direction $\nn$. Assuming (without loss of generality) $\nn=\uex$, the main contribution to $\avg{\mu_\Psi(\uex)}_\text{cnf}$ in the limit $k_0L\gg 1$ comes from a small region around $\phi=0$ and $\theta=\pi/2$.
In the plane of the ensemble ($\theta=\pi/2$), the emission is mostly confined to the interval $\Delta\phi=4\pi/k_0L$ corresponding to the distance between the first two zeros of $\sin^2[k_0L(\sin\phi)/2]/[k_0L(\sin\phi)/2]^2$. Similarly, emission in the plane perpendicular to the ensemble is collimated within an interval $\Delta\theta = 2\sqrt{4\pi/k_0L}$. 
Accordingly, we have
\be
\begin{split}
\avg{\mu_\Psi(\uex)}_\text{cnf} \simeq& \frac{N^2\Gamma_0}{16\pi}\int^{\frac{\Delta\phi}{2}}_{-\frac{\Delta\phi}{2}}\!\!\!\text{d}\phi\,\int^{\frac{\Delta\theta}{2}}_{-\frac{\Delta\theta}{2}}\!\!\!\text{d}\theta\,\pare{\frac{\sin(k_0L \phi/2)}{k_0L\phi/2}}^2\\
&\times\pare{\frac{\sin(k_0L \theta^2/4)}{k_0L\theta^2/4}}^2,
\end{split}
\ee
where we assumed $\mathcal{D}(\uu)=1/4\pi$ without loss of generality, $\phi\ll1$, and $|\theta-\pi/2|\ll 1$. Defining the constants 
\be\label{eq:Constants}
\begin{split}
	C_1 \equiv& \int_{-\pi}^\pi \text{d}\phi\, \frac{\sin^2\phi}{\phi^2} \simeq 2.8363,\\
	C_2 \equiv& \int_{-\sqrt{\pi}}^{\sqrt{\pi}}\text{d}\theta\, \frac{\sin^2\theta^2}{\theta^4}\simeq 2.30775,
\end{split}
\ee
and taking a fixed atomic density $\rho_\text{2D} = N/ L^2$, $k_0L =\sqrt{N k_0^2/\rho_\text{2D}} $, we have
\be\label{eq:mu_rel_scaling}
    \avg{\mu_\Psi(\uex)}_\text{cnf} = \frac{C_1C_2}{4\pi} \frac{N^2\Gamma_0}{(k_0L)^{\frac{3}{2}}} \sim N^{1+\frac{1}{4}},
\ee
which, once substituted back into \eqnref{eq:R_psi}, proves the scaling $R_\Psi/\Gamma_0 \sim N^{1+\frac{1}{4}}$.

{\bf Measuring the scaling of $R_\star$.} Experimentally, it is thus possible to determine the scaling of $R_\star$ by preparing the state $\ket{\Psi}$, and measuring its decay rate as a function of time for different values of $N$ while keeping the ensemble density fixed. The instantaneous decay rate can be measured directly from population loss using a quantum gas microscope as recently demonstrated in Ref.~\cite{Douglas2026} for a 2D atomic array. Alternatively, $R_\star$ can be measured by detecting the emitted light directionally using a sufficiently large NA. Our results predict the direction and solid angle of the emission, providing guidance for experiments that wish to measure the dynamics through the emitted light.

A clearer experimental signature of the scaling can be obtained by comparing the instantaneous decay of $\ket{\Psi}$ with the decay of another state that does not scale as $R_\star$. For a 2D array, such a state can be simply obtained following a similar preparation scheme as for $\ket{\Psi}$ but exciting the system from a direction perpendicular to the ensemble. This excites the product state $\ket{\Phi} = [(\ket{e}+\ket{g})/\sqrt{2}]^{\otimes N}$, whose instantaneous decay rate scales as $R_{\Phi}\sim N\Gamma_0$. From these two measurements, one can then extract $R_\Psi/R_\Phi \sim N^{1/4}$.

\clearpage
\pagestyle{plain}
\onecolumngrid

\setcounter{equation}{0}
\setcounter{figure}{0}
\setcounter{table}{0}   
\setcounter{page}{1}
\renewcommand{\theequation}{S\arabic{equation}}
\renewcommand{\thefigure}{S\arabic{figure}}
\renewcommand{\bibnumfmt}[1]{[S#1]}
\renewcommand{\citenumfont}[1]{S#1}

\begin{center}
\vspace{1cm}
\textbf{\large Supplemental Material}\\[.2cm]

Cosimo C. Rusconi$^{1,2}$, Eric Sierra$^{2}$, Wai-Keong Mok$^{3}$, Avishi Poddar$^{4}$, Simon B. J\"ager$^{5}$, Ana Asenjo-Garcia$^{2}$
{\small \itshape
${}^1$Instituto de F\'isica Fundamental - Consejo Superior de Investigaciones Cient\'ifica (CSIC), Madrid, Espa\~na.\\
${}^2$ Department of Physics, Columbia University, New York, New York 10027, USA.\\
${}^3$Institute for Quantum Information and Matter, California Institute of Technology, Pasadena, CA 91125, USA.\\
${}^4$Department of Physics, Harvard University, Cambridge, Massachusetts 02138, USA.\\
${}^5$Physikalisches Institut, University of Bonn, Nussallee 12, 53115 Bonn, Germany.
}

\newcommand{\beginsupplement}{%
        \setcounter{table}{0}
        \renewcommand{\thetable}{S\arabic{table}}%
        \setcounter{figure}{0}
        \renewcommand{\thefigure}{S\arabic{figure}}%
        \setcounter{figure}{0}
        \renewcommand{\thefigure}{S\arabic{figure}}
     }
\vspace{0.8 in}
\newcommand{\D}{\Delta}
\newcommand{\tD}{\tilde{\Delta}}
\newcommand{\K}{K_{PP}}
\newcommand{\bn}{\bar{n}_P}
\newcommand{\G}{\Gamma}
\newcommand{\LH}{\underset{L}{H}}
\newcommand{\HL}{\underset{H}{L}}
\vspace{-1in}
\end{center}

\tableofcontents

\section{Numerical results on the scaling of $\Gamma_\text{max}$}\label{sec:Gamma_max_numerics}

In this section we provide additional details on the derivation of the scaling of $\Gamma_\text{max}$ presented in \figref{fig:Fig2}. In particular we discuss how we performed the numerical simulations and the fit to obtain the scaling.

The scaling exponent $\alpha$ shown in~\figref{fig:Fig2} has been obtained by numerically solving the eigenvalue problem 
\be\label{eqS:Eigenvalue_Problem}
	\sum_j\Gamma_{ij}\psi_n^j = \sum_j\Gamma_0\frac{\sin(k_0|\rr_i-\rr_j|)}{k_0|\rr_i-\rr_j|}\psi_n^j = \Gamma_n \psi_n^i.
\ee
For arrays, the position $\rr_j$ of atom $j$ is obtained as $j_xd$ in 1D, $d(j_x,j_y)$ in 2D, and $d(j_x,j_y,j_z)$ in 3D where $j_\alpha=1\ldots N$ and $d$ is the lattice spacing. For disordered clouds, we consider atoms uniformly distributed in a $D$ dimensional box of average edge length $L=N_\text{1D} d$, where $d=\rho^{-1/D}$ is the average interatomic distance and $\rho\equiv (N_\text{1D}/L)^D \equiv N/L^D$ is the atomic density. In \appref{app:Generalizations}, we argue that the results are independent of the details of the atomic distribution by comparing to the case of atoms with a Gaussian spatial distribution.

The dependence of scaling exponent $\alpha$ on $d/\lambda_0$ is obtained as follows. For each value of $d/\lambda_0$ we numerically solve \eqnref{eqS:Eigenvalue_Problem} using the iterative eigensolver in MATLAB for a set of $q$ values for $N_\text{1D}$. We label these values as $N_\xi$ with $\xi=1,\ldots, q$. This procedure yields a set $\{\Gamma_\xi\}$ of values for $\Gamma_\text{max}$ which, after averaging over different realization of atomic positions, we fit according to the dependence $\Gamma_\text{max}=\Gamma_0\beta N^\alpha$ to extract $\alpha$. We then repeat for each value of $d/\lambda_0$. 

\subsection{Goodness of the fit and distribution of $\Gamma_\text{max}$}

\begin{figure}
	\includegraphics[width=\columnwidth]{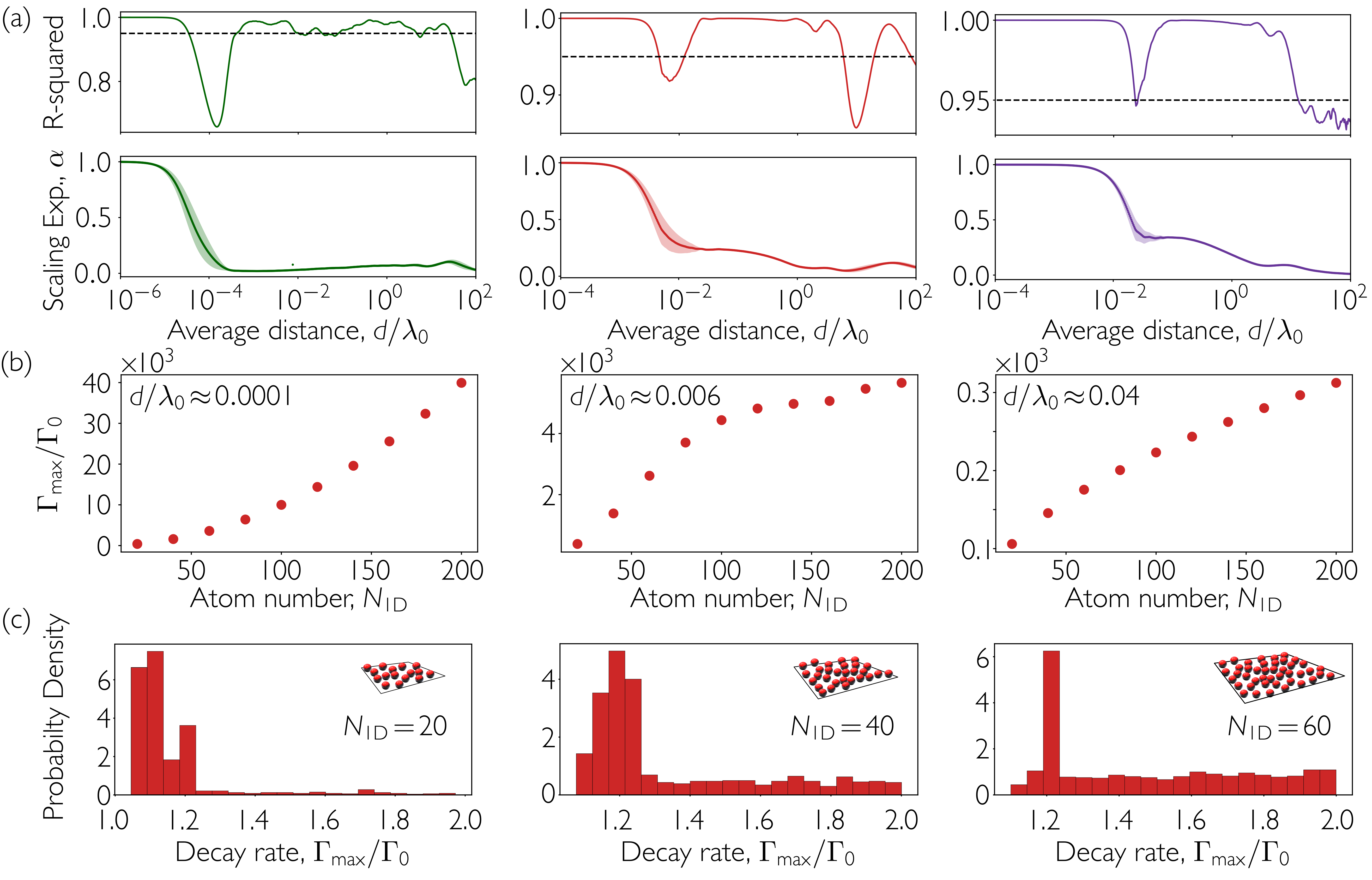}
	\caption{(a) Plot of the coefficient of determination $R$-squared parameter (top panel), as a function of $d/\lambda_0$ that quantifies the goodness of the fit for $\alpha$ (bottom panel) for uniformly distributed atoms in 1D (left), 2D (center) and 3D (right) cube. 
    (b) Plot of the dependence of $\Gamma_\text{max}/\Gamma_0$ on the number of atoms $N_\text{1D}$ for a 2D cloud. In each panel average distance is fixed and corresponds to the Dicke scaling (left panel), the dimensional scaling (right panel), and the transition region between the two (central panel).
    (c) Histogram showing the distribution of $\Gamma_\text{max}$ for a 2D cloud with $N=400$ (left), 1600 (center), and 3600 (right) with $d/\lambda_0 = 100$. The histogram is obtained from 1000 sampled values of $\Gamma_\text{max}$ and represented using 20 bins. The histogram has been normalized to have unit area.} \label{fig:FigS1}
\end{figure}

The goodness of the fit is quantified by the coefficient of determination $R^2$, which satisfies $R^2\leq 1$. For $R^2=1$ the fit is accurate, while lower values indicate less accuracy. We consider the fit to be accurate if $R^2\geq 0.95$.
In \figref{fig:FigS1}.(a), we plot the $R^2$ parameter as a function of the average atomic distance $d/\lambda_0$ for uniformly distributed atoms in 1D, 2D, and 3D cubes.
The comparison with the dependence of the scaling exponent $\alpha$ on $d/\lambda_0$ (reprinted in the bottom panel of \figref{fig:FigS1} for easier comparison) reveals that the fit is accurate in the dimensional scaling region where $\Gamma_\text{max}$ scales as ($D=1,2,3$)
\be\label{eq:Gamma_max_scaling}
    \frac{\Gamma_\text{max}}{\Gamma_0} \sim N^{\frac{1}{2}-\frac{1}{2D}}.
\ee
The fit is also accurate in the Dicke limit $d/\lambda_0 \ll 1$, where the scaling is seen to converge to $\Gamma_\text{max}\sim N\Gamma_0$ as expected from a permutational invariant system with $\Gamma_{ij}=\Gamma_0$ $\forall i,j$.

The transition region between the Dicke limit and the dimensional scaling regime occurs when the system size is comparable to the resonant wavelength, $L\sim\lambda_0$. As a function of $d/\lambda_0$, this region thus strongly depends on $N$ leading to a large error in the fit and a lower coefficient of determination (see also discussion in sec.~\ref{sec:Fit_Short_Interval}). In Fig.~\ref{fig:FigS1}(b), we plot the dependence of $\Gamma_\text{max}$ as a function of $N_\text{1D}$. While in the Dicke and dimensional scaling limits $\Gamma_\text{max}$ follows a clear power law dependence on $N_\text{1D}$, in the transition region this dependence is unclear. This results confirms that this transition regime depends sensibly on the total size of the system.

In the limit of large average interatomic separation, $d/\lambda_0 \gg 1$, $\Gamma_{ij} \rightarrow \Gamma_0 \delta_{ij}$, and thus we expect that $\alpha\rightarrow 0$.
However, in the range plotted in \figref{fig:FigS1}.(a), $\alpha\neq 0$. This is particularly evident in 1D and 2D. As indicated by the lowered $R^2$ coefficient, this scaling is a result of a poor quality of the fit model $\Gamma_\text{max}=\beta N^\alpha\Gamma_0$, which can be explained as follows.
At large \emph{average} interatomic separation, there is still a non-zero probability for two (or more) atoms to fall within a distance $d<\lambda_0$. When this happens the maximum decay rate is enhanced up to a multiple of $\Gamma_\text{max}$ for $d/\lambda_0\ll1$, corresponding to the single-excitation superradiant decay of that small cluster of closely packed atoms. It turns out that for $d\gg \lambda_0$, these clusters are made of pairs of atoms. This leads to a spreading of the probability distribution of $\Gamma_\text{max}$ across the interval $\Gamma_0\leq \Gamma_\text{max}<2\Gamma_0$ with a distribution that, for fixed density, depends strongly on $N$, with larger values of $\Gamma_\text{max}$ becoming more probable as $N$ grows [\figref{fig:FigS1}.(c)]. We also see that deviation from the limit $\alpha\rightarrow 0$ for $d/\lambda_0\gg 1$ is more pronounced for lower dimensional systems [compare lower panels in \figref{fig:FigS1}.(a)]. 
Indeed, as we shall see below, the probability for the distance $d_{ij}$ between two atoms $i$ and $j$ being smaller than a critical distance $d_c$ is proportional to 
\be\label{eq:Prob_Distnace_D}
	P_D(d_{ij}<d_c) \propto N^2\pare{\frac{d_c}{L}}^D = N\pare{\frac{d_c}{d}}^D,
\ee
valid in the limit $d_c/d\ll N^{-1/D}$. Here $d$ is the average interatomic distance and $D=1,2,3$ for 1D, 2D, and 3D arrays.
Intuitively, we expect that for uniformly distributed points it is more probable for two points to fall closer than a fixed distance $d_c$ in a lower dimensional space. We also see that \eqnref{eq:Prob_Distnace_D} increases for growing $N$, confirming the behavior observed in \figref{fig:FigS1}.(c) for the case of 2D arrays.

Let us derive the result in \eqnref{eq:Prob_Distnace_D}. 
We consider first the  1D case. Let $x_1,\cdots,x_N$ be $N$ randomly distributed points on the interval $[0,L]$. This defines $N-1$ nearest-neighbor distances $d_1,\cdots, d_{N-1}$ once the points are ordered. Let us also include the two edges such that $d_0= \text{min } x_i$ and $d_N=L-\text{max } x_i$. These variables define a simplex subject to $d_i\geq 0$ and $\sum d_i = L$ which has volume $aL^N$, where $a$ is a proportionality constant that depends on $N$. Configurations with all nearest-neighbor distances larger than a critical distance $0\leq d_c \leq L/(N-1)$ are subject to the extra constraints $d_1,\cdots,d_{N-1}\geq d_c$, which reduces the simplex volume to $a(L-(N-1)d_c)^N$, where $a$ is the same constant as before. Then, the probability of all distances being larger than $d_c$ is given by the ratio of the volumes
\begin{equation}
    P_\text{1D}(d_1\geq d_c, \cdots, d_{N-1}\geq d_c)=\left(1-\frac{(N-1)d_c}{L}\right)^N 
\end{equation}
so that
\begin{equation}\label{eq:P_1D_min_distance}
    P_\text{1D}(d_\text{min}<d_c)=1-\left(1-\frac{(N-1)d_c}{L}\right)^N.
\end{equation}
In the large $N$ limit, the second term in \eqnref{eq:P_1D_min_distance} is approximated by an exponential distribution with typical nearest neighbor distance $d_\text{mean}=L/N(N-1)$, i.e.,
\begin{equation}\label{eq:P1D_approx_exponential}
    P_\text{1D}(d_\text{min}>d_c)\to e^{-N^2d_c/L}.
\end{equation}
Using \eqnref{eq:P_1D_min_distance} and \eqref{eq:P1D_approx_exponential}, the probability of rare events when two points fall below a really small critical distance $d_c\ll d_\text{mean}$, thus reads
\begin{equation}
    P_\text{1D}(d_\text{min}<d_c\ll d_\text{mean})\approx \frac{N(N-1)d_c}{L}.
\end{equation}
In the limit $N\gg 1$, this reduces to the case $D=1$ in \eqnref{eq:Prob_Distnace_D}.

The 2D and 3D cases are not as simple, as points cannot be consistently ordered while keeping distance relations and forming a partition of the space. One can instead argue the $D$-dimensional case heuristically. Let $x_1,\cdots,x_N$ be $N$ randomly distributed points on $[0,L]^D$. Given a pair of points, the probability of them being closer than $d_c$ is roughly the ratio between the volume of a $D$-dimensional ball of radius $d_c$ and the volume of the space, so $\pi^{D/2}d_c^D/L^D\Gamma(D/2+1)$, where $\Gamma(x)$ is the Euler Gamma function . As there are $\sim N^2/2$ pairs, the expected number of pairs closer than $d_c$ is roughly $m\approx N^2\pi^{D/2}d_c^D/2L^D\Gamma(D/2+1)$. Rare events (with $d_c\ll d_\text{mean}$) are almost independent, so the number of really close pairs when $N$ goes to infinity tends to a Poisson distribution with mean $m$ (this is known as the Stein-Chen Poisson approximation \cite{S_Penrose2003}). Therefore, the probability that all pairs are at distance larger than $d_c$ satisfies
\begin{equation}
    P_D(d_\text{min}>d_c) \to e^{-\alpha_DN^2d_c^D/L^D},
\end{equation}
where $2\alpha_D=\pi^{D/2}/\Gamma(D/2+1)$ is the volume of a $D$-dimensional unit ball. Therefore, really close pairs happen with a probability
\begin{equation}
    P_{D}(d_\text{min}<d_c\ll d_\text{mean})\approx \alpha_D N^2\left(\frac{d_c}{L}\right)^D = \alpha_D N \left(\frac{d_c}{d}\right)^D,
\end{equation}
meaning that the probability that there are two points falling really close decreases with dimensionality (when $d_c\ll d_\text{mean} < d$). This concludes our proof of \eqnref{eq:Prob_Distnace_D}.

\subsection{Beyond the scalar model: tensorial Green's function and non-uniform density}\label{app:Generalizations}

We now present numerical evidence that the scaling law in \eqnref{eq:Gamma_max_scaling} holds beyond the particular case of scalar dissipative interaction matrix in \eqnref{eqS:Eigenvalue_Problem} and uniform atomic distribution.

\begin{figure}
	\includegraphics[width=\columnwidth]{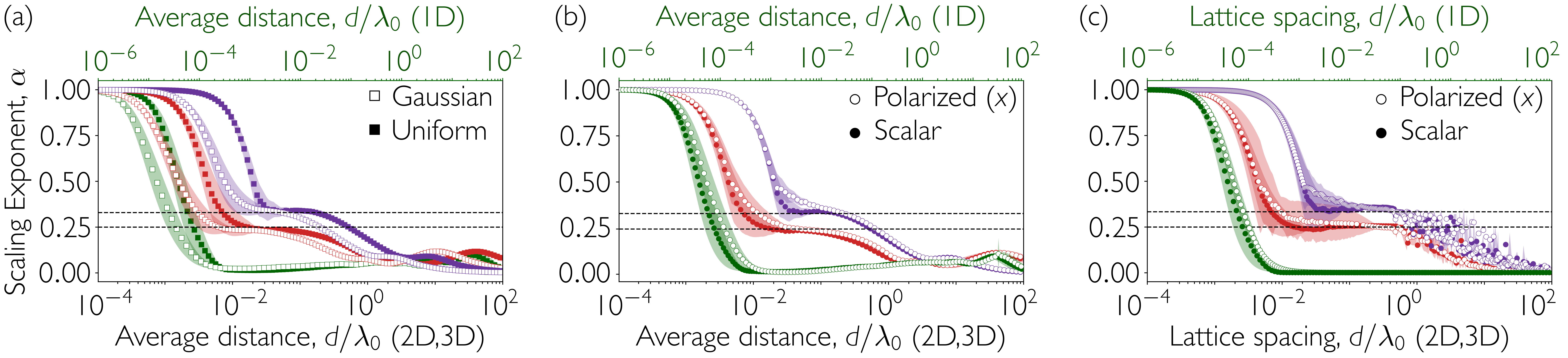}
	\caption{Effects of different atomic distribution and tensorial Green's function on the scaling of $\Gamma_\text{max}$ (obtained from a best fit to $\Gamma_\text{max}=\beta N^\alpha\Gamma_0$). (a) Comparison of the scaling exponent $\alpha$ between a disordered cloud with uniform (filled squares) and Gaussian (empty squares) atomic distributions. (b) Comparison of the scaling exponent $\alpha$ in atomic clouds for the scalar model (filled circles) and the tensorial model in \eqnref{eq:green} with atoms polarized along the $x$-axis (empty circles). (c) Same as (b) but for ordered arrays. In all panels, we show the case of atoms in a line (green markers), square (red markers) or cube (purple markers) with linear dimensions $L=N_\text{1D}d$.}\label{fig:Fig_S2}
\end{figure}

First we consider the effect of different atomic density distributions. We assume a Gaussian density profile,
\be\label{eq:Gaussian_Density}
	\rho_D(\rr) = N\frac{e^{-\rr^2/2L^2}}{(2\pi L^2)^{D/2}}.
\ee
In \figref{fig:Fig_S2}.(a), we plot the scaling of $\alpha$ for the case of uniform and Gaussian atomic densities. Qualitatively the two curves follow a similar trend. In particular, for both choices the scaling of $\alpha$ approaches the value in \eqnref{eq:Gamma_max_scaling} in the same parameter region. The Gaussian distribution leads to an earlier deviation from both the Dicke scaling and the scaling in \eqnref{eq:Gamma_max_scaling}. This can be intuitively understood:  the effective size of the system for the density in \eqnref{eq:Gaussian_Density} is larger than if one assumes a uniform distribution on a $D$-cube of side $L$.

Let us now consider the effects of atomic polarization. These can be included considering the general form of the dissipative matrix elements for atoms polarized along the direction $\hat{\dpp}$,
\be\label{eq:Gamma_ij}
	\Gamma_{ij}= \frac{6\pi \Gamma_0}{k_0} \hat{\dpp}^{*}\cdot \text{Im}\,\textbf{G}(\textbf{r}_i,\textbf{r}_j,\omega_0) \cdot \hat{\dpp},
\ee
where in 3D vacuum the free space electromagnetic Green's tensor is given by~\cite{S_jackson}
\be\label{eq:green}
    \textbf{G} (\textbf{r},\w_0) \equiv \frac{e^{\im k_0 r}}{4\pi k_0^2r^3}\Big[(k_0^2r^2+\im k_0 r-1)\mathds{1}+(-k_0^2r^2-3\im k_0r+3)\frac{\textbf{r}\otimes\textbf{r}}{r^2}\Big],
\ee	
where $\mathbf{r} \equiv \mathbf{r}_i - \mathbf{r}_j$ and $r=|\rr|$. Figure~\ref{fig:Fig_S2}(b) and (c) compares explicitly the dependence of $\alpha$ for both the tensorial and scalar models of $\Gamma_{ij}$ for an atomic cloud and an array, respectively. A remarkably similar behavior is obtained for both models. This is in agreement with the results of Ref.~\cite{S_Bellando2021}, which compared the tensorial and scalar models for atomic clouds, finding similar distributions for the eigenvalues.

\begin{figure}
    \centering
    \includegraphics[width=\columnwidth]{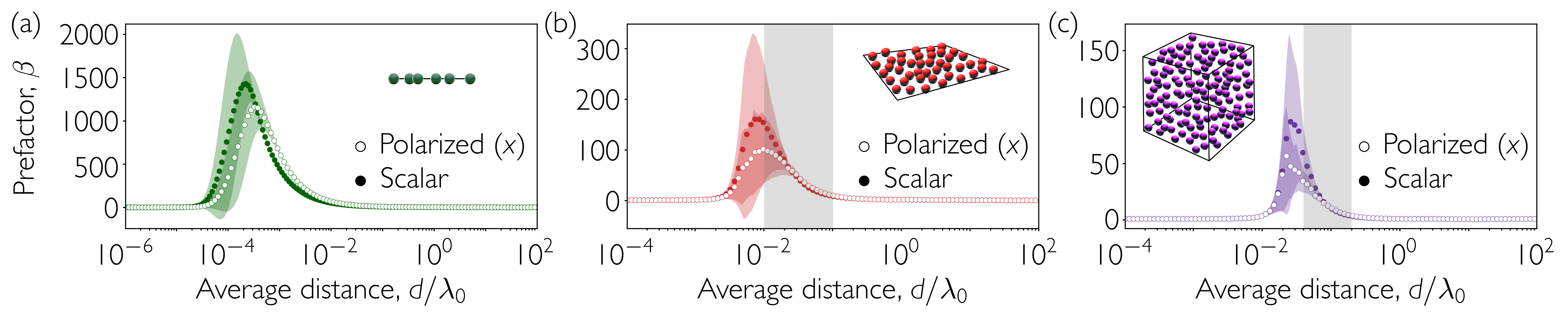}
    \caption{Plot of the prefactor $\beta$ in $\Gamma_\text{max} = \beta N^\alpha$ as a function of $d/\lambda_0$ for (a) 1D, (b) 2D, and (c) 3D ensembles of atoms. For each panel we compare the result for the scalar model (filled circles) and the tensorial model (empty circles).}
    \label{fig:Fig_beta}
\end{figure}
Finally, in \figref{fig:Fig_beta} we plot the dependence of the prefactor $\beta$ on the average interatomic distance $d/\lambda_0$ and compare the results of the scalar and the tensorial model.
We observe qualitatively similar results. 
For the case of 2D and 3D ensembles, we highlight the region where the scaling exponent $\alpha$ follows the dimensional scaling $(1/2-1/2D)$. As for the case of atomic arrays (see supplemental material of \cite{S_Mok2026}), the error in the fit for $\beta$ is larger in the transition region between the Dicke limit and the dimensional scaling. Note, however, that in the relevant region of achievable atomic densities---for largest values of $d/\lambda_0$ in the dimensional scaling regime---the relative error on $\beta$ is small, as for the case of atomic arrays.
To conclude, we remark that a difference in the prefactor $\beta$ does not affect the scaling $\alpha$ of $\Gamma_\text{max}$.

\subsection{Scaling of $\Gamma_\text{max}$ for small variation of N.} \label{sec:Fit_Short_Interval}

\begin{figure}
    \centering
    \includegraphics[width=0.6\columnwidth]{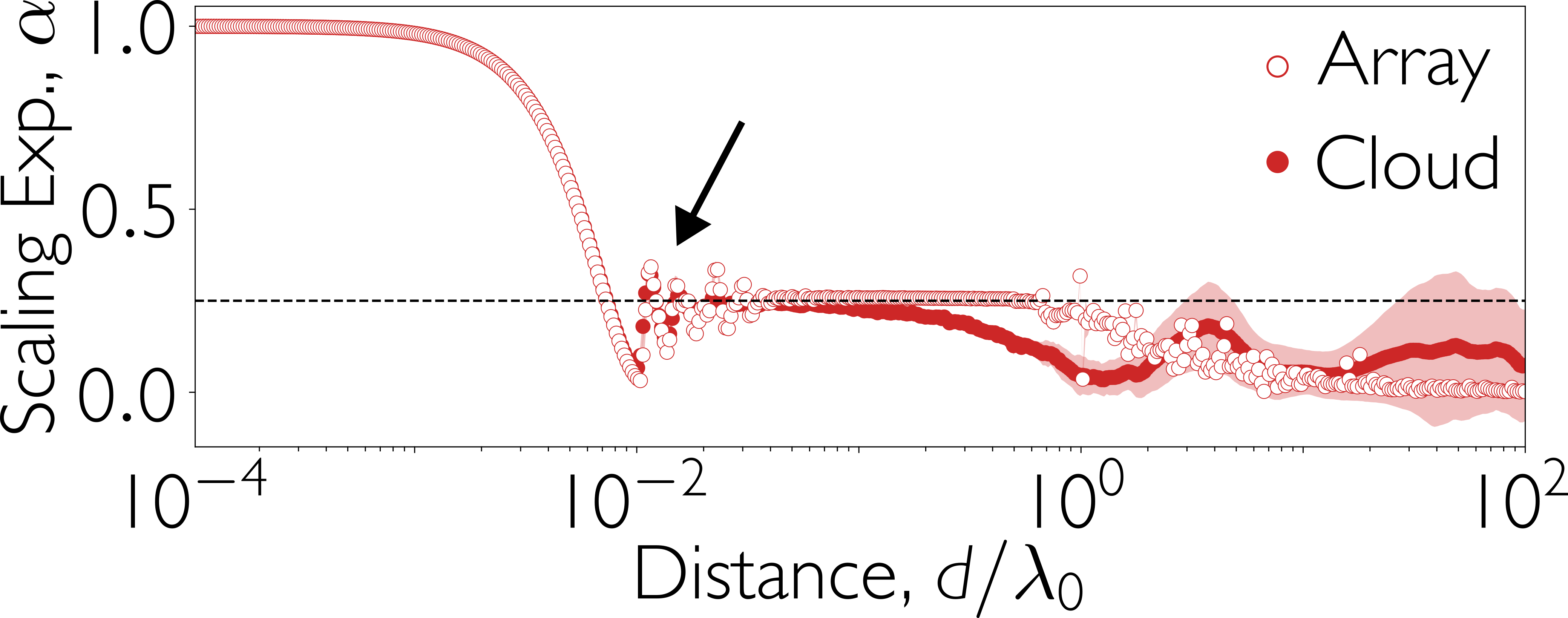}
    \caption{Scaling of $\Gamma_\text{max}$ from a fit to $\beta N^\alpha \Gamma_0$, as a function of the average interatomic separation $d/\lambda_0$ for 2D clouds and arrays. The colored region represents the $1\sigma$ confidence interval. The fits are done over a region $N_\text{1D}=90,91,\ldots,100$. For clouds the results are averaged over 100 realizations. The black arrow indicates the region where $L\sim\lambda_0$.}
    \label{fig:Fit_Local}
\end{figure}

In the main text, we mostly focused on fitting the scaling of $\Gamma_\text{max}$ over a large range of values for the system size $N$. In experiments, however, it is sometimes more common to vary $N$ over a small region around a central average value. In the following, we show that performing the fit over a small region of $N$ does not change our conclusions.

In \figref{fig:Fit_Local} we fit the scaling of $\Gamma_\text{max}$ for 2D clouds and arrays over a much smaller range of variation in the system size $N$ as compared to \figref{fig:Fig2}(a). For both arrays and clouds, the scaling coefficient displays very sharp oscillations in the region where the ensemble's linear dimension is of the order of the transition wavelength, $L\sim \lambda_0$ (black arrow in \figref{fig:Fit_Local}). These oscillations are due to a change in the mode structure of $\boldsymbol{\psi}_\text{max}$. For $L\ll \lambda_0$, $\Gamma_{ij}>0$ $\forall i,j$. Accordingly, constructive interference occurs when atoms oscillate parallel to each other. As $d/\lambda_0$ increases, $\Gamma_{ij}$ can take negative signs between distant pairs of atoms. For $\Gamma_{ij}<0$, constructive interference in the emission between atoms $i$ and $j$ occurs when they oscillate with opposite phase (anti-parallel). The brightest mode thus acquires a spatial phase modulation across the system to maximize constructive interference. 
The oscillations are eventually damped as for large $d/\lambda_0$ the mode structure of $\boldsymbol{\psi}_\text{max}$ becomes very susceptible to variations in $N$.

We remark that when fitting over a large span for $N$, as done in Figs.~\ref{fig:Fig2} and \ref{fig:Fig_S2}, these oscillations are averaged out as the value of $d/\lambda_0$ at which $L\sim \lambda_0$ depends sensitively on $N$.

\section{Analytical scaling of $\Gamma_\text{max}$ for dense atomic clouds}\label{sec:Scaling_Gamma_max_Clouds}

In this section we compute the largest eigenvalue of $\boldsymbol{\Gamma}$ for a dense atomic cloud. In the large density limit, the spectrum of $\boldsymbol{\Gamma}$ in \eqnref{eqS:Eigenvalue_Problem} is asymptotically close to the spectrum of the integral operator~\cite{S_Koltchinskii2000, S_Williams2000},
\be\label{eqS:Fredholm_Eq}
	\int_{V_D}\!\!\text{d}\rr'\, \rho_D\frac{\sin(k_0 |\rr-\rr'|)}{k_0|\rr-\rr'|} \psi_\mu(\rr') = \frac{\Gamma_\mu}{\Gamma_0}\psi_\mu(\rr).
\ee
The integral is computed over a $D$-dimensional volume $V_D$ corresponding to a line of length $L$, a disk of radius $L$, and a sphere of radius $L$ for ensembles in 1D, 2D, and 3D, respectively. In the following, we consider atomic ensembles with uniform densities, that is $\rho_\text{1D}=N/L$, $\rho_\text{2D}=N/\pi L^2$, and $\rho_\text{3D}=3 N/4\pi L^3$.

We compute the largest eigenvalue of \eqnref{eqS:Fredholm_Eq} using Gelfand's formula. A matching lower bound follows from the eigenfunction ansatz in Eq.~\eqref{eq:Ansatz_Eigenstates}, which yields eigenvalues with the same scaling as $\Gamma_{\max}$ -- since any eigenvalue satisfies $\Gamma_n \leq \Gamma_{\max}$ by definition, these eigenvalues certify that the upper bound is tight. Together, the two approaches establish the scaling of $\Gamma_{\max}$ with system size. Noting that $\Tr[\boldsymbol{\Gamma}^m]= \sum_\mu \Gamma_\mu^m$, we write the inequality
\be\label{eq:Trace_m}
    1 \leq \frac{\sqrt[m]{\Tr[\Gamma^m]}}{\Gamma_\text{max}}\leq N^{1/m},
\ee
where $\Gamma$ is the Kernel of the integral operator as defined in \eqnref{eq:Gamma_kernel_scalar}. From Eq.~\eqref{eq:Trace_m} we obtain the largest eigenvalue as 
\be\label{eq:Gamma_max_limit_trace}
    \frac{\Gamma_\text{max}}{\Gamma_0} = \lim_{m\rightarrow\infty} \frac{\sqrt[m]{\Tr[\Gamma^m]}}{\Gamma_0}= \lim_{m\rightarrow \infty} \spare{\int\!\!\text{d}\rr_1 \ldots \text{d}\rr_m \,\rho_D^m \frac{\sin(k_0 |\rr_1-\rr_2|)}{k_0|\rr_1-\rr_2|}\ldots \frac{\sin(k_0 |\rr_m-\rr_1|)}{k_0|\rr_m-\rr_1|}}^{\frac{1}{m}}.
\ee
We show below that this yields a scaling consistent with the results of the numerical simulation presented in \figref{fig:Fig2} in the main text and \figref{fig:Fig_SDP} in Sec.~\ref{sec:SDP} below. This approach was first used by Ressayre and Tallet~\cite{S_Ressayre1976, S_Ressayre1977} for a cylinder-shaped cloud of atoms. 
Though they do not explicitly discuss the scaling of $\Gamma_\text{max}$ with the number of atoms $N$, our results for 3D ensembles are consistent with theirs. Our results, however, differ for a disk, as Ressayre and Tallet performed an approximation leading to a looser bound on $\Gamma_\text{max}$. We now consider $D=1,2,3$ separately.

\subsection{1D Clouds}
To obtain the scaling of the largest emission rate for a 1D cloud from \eqnref{eq:Gamma_max_limit_trace}, we need to compute the integral
\be\label{eq:Lambda_1D}
\begin{split}
    T_m^\text{1D} \equiv \frac{\Tr[\Gamma^m]}{\Gamma_0^m} =& \int_{-L/2}^{L/2}\!\!\text{d}x_1\ldots \text{d}x_m \,\rho_\text{1D}^m \frac{\sin[k_0 (x_1-x_2)]}{k_0(x_1-x_2)}\ldots \frac{\sin[k_0 (x_m-x_1)]}{k_0(x_m-x_1)}.
\end{split}
\ee
For greater clarity, it is useful to consider the particular cases of $m=2$ and $m=3$ first and later generalize to any value of $m$.

{\bf Case $m=2$.} For this case the integral in \eqnref{eq:Lambda_1D} takes the form
\be\label{eq:Lambda1D_m=2}
\begin{split}
T^\text{1D}_2 =& \rho_\text{1D}^2 \int_{-L/2}^{L/2}\!\!\text{d}x \text{d}y \pare{\frac{\sin[k_0(x-y)]}{k_0(x-y)}}^2 = \rho_\text{1D}^2\int_{\mathbb{R}}\!\!\text{d}x \text{d}y \pare{\frac{\sin[k_0(x-y)]}{k_0(x-y)}}^2\boldsymbol{1}_L(x)\boldsymbol{1}_L(y)\\
=& \rho_\text{1D}^2\int_{\mathbb{R}}\!\!\text{d}w \text{d}u \pare{\frac{\sin(k_0w)}{k_0w}}^2\boldsymbol{1}_L(w)\boldsymbol{1}_L(w+u)
\end{split}
\ee
where we introduce the indicator function of the interval $[-L/2,L/2]$, i.e.,
\be\label{eq:indicator_function}
	\boldsymbol{1}_L(x) =
	\left\{
	\begin{array}{ll}
	1& \quad \text{$-L/2<x<L/2$},\\
	0& \quad \text{otherwise},\\
	\end{array}
	\right .
\ee 
and in the last step we made the change of variables $w=x-y$ and $u=y$. This allows us to compute separately the integral over $u$, which is the overlap of two identical indicator functions, one shifted by a fixed quantity $w$. The integral is straightforward and yields
\be\label{eq:Integral_two_indicator_function}
    \int_{\mathbb{R}}\!\!\text{d}u\,\boldsymbol{1}_L(w)\boldsymbol{1}_L(w+u) =\left\{
	\begin{array}{ll}
	L-|w|& \quad \text{$-L<w<L$},\\
	0& \quad \text{otherwise}.\\
	\end{array}
	\right .
\ee
It can be upper bounded as
\be\label{eq:Upper_Bound_Indicator_Integral}
    \int_{\mathbb{R}}\!\!\text{d}u\,\boldsymbol{1}_L(w)\boldsymbol{1}_L(w+u)\leq L \boldsymbol{1}_{2L}(w).
\ee
Substituting this result into \eqnref{eq:Lambda1D_m=2} and using that 
$\sin(k_0w)/k_0w = (2k_0)^{-1}\int_\mathbb{R}\! \text{d}q\, e^{-\im q w}\boldsymbol{1}_{2k_0}(q)$, we obtain
\be\label{eq:Calcualtion_m=2}
\begin{split}
    T_2^{\text{1D}} \leq& L\frac{\rho_\text{1D}^2}{(2k_0)^2}\int_{\mathbb{R}}\!\!\text{d}w\, \boldsymbol{1}_{2L}(w) \int_{\mathbb{R}}\!\! \text{d}q_1 \text{d}q_2\, \boldsymbol{1}_{2k_0}(q_1)\boldsymbol{1}_{2k_0}(q_2)e^{-\im (q_1+q_2) w}\\
    = & L\frac{\rho_\text{1D}^2}{2k_0^2}\int_{\mathbb{R}}\! \text{d}q_1 \text{d}q_2\,  \frac{\sin[(q_1+q_2)L]}{(q_1+q_2)}\boldsymbol{1}_{2k_0}(q_1)\boldsymbol{1}_{2k_0}(q_2)= L\frac{\rho_\text{1D}^2}{2k_0^2} \int_{\mathbb{R}}\! \text{d}v \text{d}s\, \frac{\sin(vL)}{v} \boldsymbol{1}_{2k_0}(s)\boldsymbol{1}_{2k_0}(v-s),
\end{split}
\ee
where in the last step we made the change of variables $v=q_1+q_2$ and $s=q_2$. The integral over $s$ is the same as \eqnref{eq:Integral_two_indicator_function} with obvious changes in the interval of the indicator function and it can be similarly upper bounded as in \eqnref{eq:Upper_Bound_Indicator_Integral}. This leads to the result
\be\label{eq:Lambda_m=2_final}
    T_2^{\text{1D}} \leq L\frac{\rho_\text{1D}^2}{k_0} \int_{\mathbb{R}}\!\! \text{d}v\, \frac{\sin(vL)}{v}\boldsymbol{1}_{4k_0}(v) = 2L\frac{\rho^2_\text{1D}}{k_0} \text{Si}(2k_0L),
\ee
where we defined the Sine Integral function $\text{Si}(x)=\int_0^x \text{d}t \sin(t)/t$.

{\bf Case $m=3$.} This case is solved following an analogous procedure. It is, however, instructive to look at it in some detail. Performing a similar change of variables as the one described in \eqnref{eq:Lambda1D_m=2}, we have
\be\label{eq:Lambda_m=3}
    T_3^{\text{1D}} = \rho_\text{1D}^3\int_{\mathbb{R}}\!\!\text{d}w_1 \text{d}w_2 \text{d}u \frac{\sin(k_0w_1)}{k_0w_1}\frac{\sin(k_0w_2)}{k_0w_2}\frac{\sin[k_0(w_1+w_2)]}{k_0(w_1+w_2)} \boldsymbol{1}_L(u)\boldsymbol{1}_L(u+w_2)\boldsymbol{1}_L(u+w_1+w_2).
\ee
We now proceed to compute the integral over $u$ that involves only the indicator functions in \eqnref{eq:Lambda_m=3}. The integral is non-zero only on a region defined by the conditions $-L\leq w_1\leq L$ and $-L\leq w_1+w_2\leq L$, and it can thus be upper bounded as
\be\label{eq:Integral_Indicator_m=3}
    \int_{\mathbb{R}}\!\!\text{d}u\,\boldsymbol{1}_L(u)\boldsymbol{1}_L(u+w_2)\boldsymbol{1}_L(u+w_1+w_2) \leq L\,\boldsymbol{1}_{2L}(w_1)\boldsymbol{1}_{2L}(w_2).
\ee
The indicator functions on the right-hand side of \eqnref{eq:Integral_Indicator_m=3} define a region $-L\leq w_1,w_2\leq L$, which is larger than the actual support of the integral as discussed above. Substituting \eqnref{eq:Integral_Indicator_m=3} into \eqnref{eq:Lambda_m=3} and following the same procedure as in \eqnref{eq:Calcualtion_m=2} we obtain
\be
\begin{split}
    T_3^{\text{1D}} =& \frac{L \rho_\text{1D}^3}{2k_0^3}\int_\mathbb{R}\!\!\text{d}v_1\text{d}v_2\, \frac{\sin(v_1 L)}{v_1}\frac{\sin(v_2 L)}{v_2}\int_\mathbb{R}\!\text{d}s\,\boldsymbol{1}_{2k_0}(s)\boldsymbol{1}_{2k_0}(s-v_1)\boldsymbol{1}_{2k_0}(s-v_2)\leq  \frac{L\rho_\text{1D}^3}{k_0^2} [2\text{Si}(2k_0L)]^2.
\end{split}
\ee
In the last step we used \eqnref{eq:Integral_Indicator_m=3} with the substitution $L\rightarrow 2k_0$.

{\bf Generic value $m$.} The procedure illustrated above for $m=2$ and $m=3$ can be easily generalized to the case of a generic value of $m$. Defining $w_1\equiv x_1-x_2, \ldots, w_{m-1}\equiv x_{m-1}-x_m$, and $u=x_m$, we have 
\be\label{eq:Lambda_1D_m_general}
\begin{split}
    T_m^\text{1D} 
    =& \rho_\text{1D}^m\int_\mathbb{R}\!\!\!\text{d}w_1\ldots \text{d}w_{m-1} \, \frac{\sin k_0 w_1}{k_0w_1}\ldots \frac{\sin[k_0 (w_1+\ldots+w_{m-1})]}{k_0(w_1+\ldots+w_{m-1})} \int_\mathbb{R}\!\!\text{d}u\,\boldsymbol{1}_L(u)\boldsymbol{1}_L(u+w_{m-1})\ldots \boldsymbol{1}_L(u+w_{m-1}+\ldots w_1)\\
    \leq& \frac{L\rho_\text{1D}^m}{(2k_0)^m}\! \int_\mathbb{R}\!\!\!\text{d}w_1\ldots \text{d}w_{m-1} \,\boldsymbol{1}_L(w_1)\ldots \boldsymbol{1}_L(w_{m-1}) \int_\mathbb{R}\!\!\text{d}q_1\ldots \text{d}q_m\, \boldsymbol{1}_{2k_0}(q_1)e^{\im q_1 w_1}\ldots \boldsymbol{1}_{2k_0}(q_m)e^{\im q_m(w_1+\ldots+\w_{m-1})}\\
    =& \frac{L\rho_\text{1D}^m}{2 k_0^m} \int_\mathbb{R}\!\! \text{d}q_1\ldots \text{d}q_m\, \frac{\sin[(q_1+q_m)L]}{q_1+q_m}\ldots \frac{\sin[(q_{m-1}+q_m)L]}{q_{m-1}+q_m}\boldsymbol{1}_{2k_0}(q_1)\ldots\boldsymbol{1}_{2k_0}(q_m)\\
    \leq & \frac{L \rho_\text{1D}^m}{k_0^{m-1}} \spare{2\text{Si}(2k_0L)}^{m-1} 
\end{split}
\ee
In the first and last steps, we straightforwardly generalized the result in \eqnref{eq:Integral_Indicator_m=3} to upper bound the integral over a product of $m$ indicator functions of the interval $[-L/2,L/2]$ or $[-k_0,k_0]$ respectively.

We now substitute the result of \eqnref{eq:Lambda_1D_m_general} into  \eqnref{eq:Gamma_max_limit_trace} and obtain the following upper bound on the largest eigenvalue of \eqnref{eqS:Fredholm_Eq} in one dimension
\be
    \frac{\Gamma_\text{max}}{\Gamma_0} = \lim_{m\rightarrow \infty}\sqrt[m]{T_m^\text{1D}}\leq \lim_{m\rightarrow \infty} \frac{L^{1/m}\rho_\text{1D}}{k_0^{1-1/m}} \spare{2\text{Si}(2k_0L)}^{1-1/m} = \frac{2\rho_\text{1D}}{k_0}\text{Si}(2k_0L) \simeq \frac{\pi\rho_\text{1D}}{k_0} = O(1),
\ee
where we used that $2\text{Si}(2k_0L)\simeq \pi$ in the limit of large ensembles  ($k_0L\gg 1$).

\subsection{2D Clouds}

To obtain the scaling of the largest emission rate for a 2D cloud from \eqnref{eq:Gamma_max_limit_trace}, we need to compute the integral
\be\label{eq:Trace_m_2D}
\begin{split}
    T_m^\text{2D} \equiv \frac{\Tr[\Gamma^m]}{\Gamma_0^m} = \rho_{2D}^m \int_{D_L} \!\!\!\text{d}\rr_1\ldots \text{d}\rr_m \frac{\sin(k_0|\rr_1-\rr_2|)}{k_0 |\rr_1-\rr_2|}\ldots \frac{\sin(k_0|\rr_{m-1}-\rr_m|)}{k_0 |\rr_{m-1}-\rr_m|}\frac{\sin(k_0|\rr_m-\rr_1|)}{k_0 |\rr_{m}-\rr_1|},
\end{split}
\ee
where $D_L$ is a disk of radius $L$. We introduce the relative coordinates $\ww_j \equiv \rr_j-\rr_{j+1}$ for $j=1,\ldots,m-1$, as well as define $\boldsymbol{1}_A(\rr)$ as the indicator function of a disk of radius $L$ centered at $\rr=0$, and rewrite \eqnref{eq:Trace_m_2D} as
\be\label{eq:T_m_2D}
\begin{split}
    T_m^\text{2D} =& \rho_{2D}^m
    \int_{\mathbb{R}^2}\!\!\text{d}\ww_1\ldots \text{d}\ww_{m-1}\text{d}\rr \, \frac{\sin(k_0 |\ww_1|)}{k_0|\ww_1|}\ldots \frac{\sin(k_0 |\ww_{m-1}|)}{k_0|\ww_{m-1}|}\frac{\sin(k_0 |\ww_1+\ldots+\ww_{m-1}|)}{k_0|\ww_{1}+\ldots+\ww_{m-1}|}\times\\
    &\times\boldsymbol{1}_A(\rr)\boldsymbol{1}_A(\rr+\ww_{m-1})\ldots \boldsymbol{1}_A(\rr+\ww_{m-1}+\ldots+\ww_1)\\
    \leq&  \pi L^2 \rho_{2D}^m\int_{D_{2L}}\!\!\text{d}\ww_1\ldots \text{d}\ww_{m-1}\, \frac{\sin(k_0 |\ww_1|)}{k_0|\ww_1|}\ldots \frac{\sin(k_0 |\ww_{m-1}|)}{k_0|\ww_{m-1}|}\frac{\sin(k_0 |\ww_1+\ldots+\ww_{m-1}|)}{k_0|\ww_{1}+\ldots+\ww_{m-1}|},
\end{split}
\ee
where we upper bounded the integral over $\rr$ using the same argument as for 1D in \eqnref{eq:Integral_Indicator_m=3}. We now proceed to rewrite the integral in Fourier space. We introduce the functions
\be\label{eq:f_k_transform}
\begin{split}
    f(\kk) \equiv& \inv{2\pi}\int_{\mathbb{R}^2}\!\!\text{d}\rr\, \frac{\sin(k_0 r)}{k_0r} e^{\im \kk\cdot \rr} = \inv{k_0} \frac{\Theta(k_0-k)}{\sqrt{k_0^2-k^2}},\\
    \Delta_{2L}(\kk) \equiv& \inv{(2\pi)^2}\int_{D_{2L}}\!\!\!\text{d} \rr\, e^{\im \kk\cdot\rr} = \frac{L}{2\pi}\frac{J_1(k L)}{k}.
\end{split}
\ee
Here, $\Theta(x)$ is the Heaviside step function, which is one for $x>0$ and zero otherwise, and $J_1(x)$ is the Bessel function of the first kind of order one. We note that $\Delta_{2L}(\kk)$ is normalized when integrated over all $k$-space.
Substituting \eqnref{eq:f_k_transform} into \eqnref{eq:T_m_2D} we obtain
\be\label{eq:T_m_2D_Fourier}
    T_m^\text{2D} \leq \pi (2\pi)^{m-1} L^2 \rho_{2D}^m \int \text{d}\kk_1\ldots \text{d}\kk_m\, f(\kk_1)\ldots f(\kk_m) \Delta_{2L}(\kk_1+\kk_m)\ldots\Delta_{2L}(\kk_{m-1}+\kk_m).
\ee
The function $\Delta_{2L}(\kk)$ is a nascent Dirac delta function, \ie $\lim_{L\rightarrow \infty} \Delta_{2L}(\kk)= \delta(\kk)$, which for a finite value of $L$ has a width $1/L$ around $\kk=0$.
In the limit of large system size ($k_0L\gg1$), it is tempting to approximate $\Delta_{2L}(\kk)\simeq \delta(\kk)$ in \eqnref{eq:T_m_2D_Fourier}. However, this yields an integral of the form $\int_0^{k_0} \text{d}\kk (k_0^2-k^2)^{-m/2}$ that diverges at $k=k_0$. This is not surprising, as $\Gamma_\text{max}$ diverges in the limit of an infinite system. Thus, it is crucial to maintain the finite width of $\Delta_{2L}(\kk)$ to determine the functional form of the divergence as $k_0L$ grows.
In particular, $\Delta_{2L}(\kk)$ can only resolve $\kk$ within a region of width $1/L$. 
This effectively smooths the divergence of $f(\kk)$ at $k=k_0$, yielding a finite result for \eqnref{eq:T_m_2D_Fourier}. 
We approximate this smoothening as $\Delta_{2L}(\kk-\kk')\approx k^{-1}\delta(k-k'-1/L)\delta(\theta-\theta')$, where we expressed the Dirac delta in polar coordinates in 2D $k$-space, and $1/L$ represents the finite resolution due to the finite size of the system. Substituting this result into \eqnref{eq:T_m_2D_Fourier} we obtain
\be\label{eq:T_m_2D_cut_off}
\begin{split}
    T_m^\text{2D} \leq& \pi (2\pi)^{m-1} L^2 \pare{\frac{\rho_{2D}}{k_0}}^m \!\int_0^{k_0}\!\!\text{d}k\, \frac{k}{\sqrt{k_0^2-k^2}}\spare{k_0^2-\pare{k-\frac{1}{L}}^2}^{-\frac{m}{2}}\\
    =& \pi (2\pi)^{m-1} k_0^2 L^2 \pare{\frac{\rho_{2D}}{k_0^2}}^m \!\!\underbrace{\int_0^1\!\! \text{d}x\, \frac{x}{\sqrt{1-x^2}}\spare{1-\pare{x-\frac{1}{k_0L}}^2}^{-\frac{m}{2}}}_{\equiv\mathcal{I}}
\end{split}
\ee
The integral in \eqnref{eq:T_m_2D_cut_off} can be solved exactly but its solution is complicated and not particularly enlightening. As we are interested in the scaling of $T_m^\text{2D}$ in the limit of large system sizes, we can easily evaluate the scaling of the integral in this limit. To this aim, we write
\be\label{eq:Integral_I_2D_scaling}
    \mathcal{I} = \int_0^{1-\Lambda} \!\!\!\text{d}x\, \frac{x}{\sqrt{1-x^2}}\spare{1-\pare{x-\frac{1}{k_0L}}^2}^{-\frac{m}{2}} + \underbrace{\int_{1-\Lambda}^{1}\!\!\! \text{d}x\, \frac{x}{\sqrt{1-x^2}}\spare{1-\pare{x-\frac{1}{k_0L}}^2}^{-\frac{m}{2}}}_{\square},
\ee
where $(k_0L)^{-1} \ll \Lambda \ll 1$ and $\Lambda$ is fixed and is not scaled with $k_0L$. The first integral on the right hand side of \eqnref{eq:Integral_I_2D_scaling} gives a finite contribution even in the limit $k_0L\rightarrow\infty$, because it does not contain the divergence at $x=1$. The dominant contribution to $T_m^\text{2D}$ thus arises from the second term in \eqnref{eq:Integral_I_2D_scaling}. 
To evaluate this contribution, we define $x=1-s$ and since $s\leq \Lambda \ll 1$ we write
\be
    \square \simeq 2^{-\frac{m+1}{2}} \int_0^\Lambda \text{d}s \frac{1}{\sqrt{s}}\pare{s+\frac{1}{k_0L}}^{-\frac{m}{2}} = \frac{(k_0L)^{(m-1)/2}}{2^{(m+1)/2}}\int_0^{\Lambda k_0L}\!\!\! \text{d}u\, u^{-1/2}(1+u)^{-m/2}.
\ee
In the limit of large system size $k_0L\gg1$, we can extend the upper limit of the integral to infinity. The resulting integral can be exactly computed in terms of the Euler Gamma function $\Gamma(x)$. We thus finally obtain
\be
    \square = \frac{\sqrt{\pi}\Gamma\pare{\frac{m-1}{2}}}{\Gamma\pare{\frac{m}{2}}}\frac{(k_0L)^{(m-1)/2}}{2^{(m+1)/2}}.
\ee
Substituting this result back into the \eqnref{eq:T_m_2D_cut_off}, we finally obtain for the largest eigenvalue of $\boldsymbol{\Gamma}$
\be
    \frac{\Gamma_\text{max}}{\Gamma_0} = \lim_{m\rightarrow \infty}\sqrt[m]{T_m^{2D}} \sim \sqrt{k_0L}\sim N^{1/4}.
\ee

\subsection{3D Clouds}

\begin{figure}
    \centering
    \includegraphics[width=0.5\columnwidth]{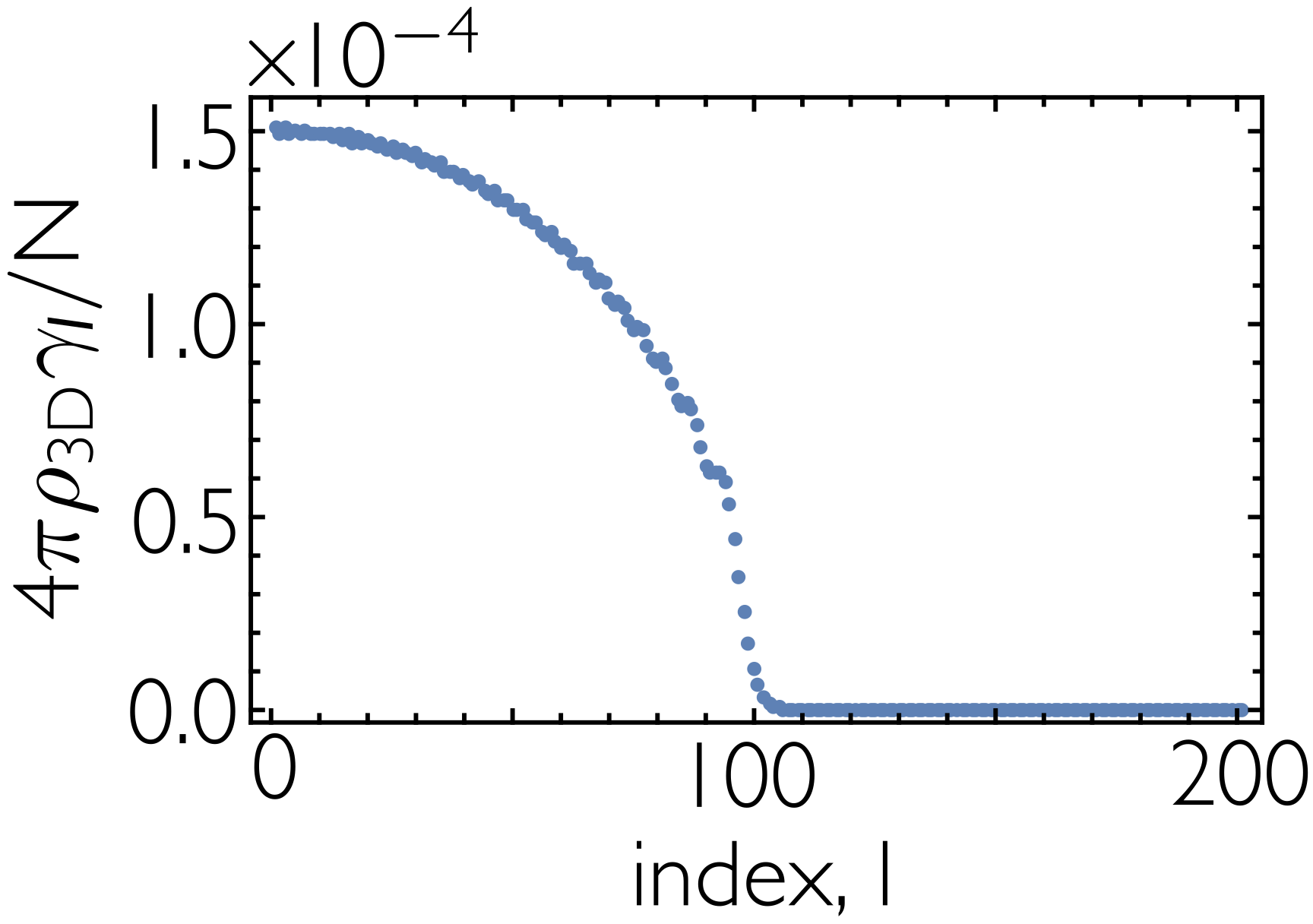}
    \caption{Plot of $\gamma_l$ as a function of $l$ for $k_0L=100$.}
    \label{fig:Plot_Gamma_l_3D}
\end{figure}

Analogously to the previous cases of 1D and 2D clouds, we need to evaluate the integral
\be\label{eq:Lambda_3D}
\begin{split}
    T_m^\text{3D} \equiv&  \int_{S_L}\!\!\text{d}\rr_1\ldots \text{d}\rr_m \, \rho_\text{3D}^m\frac{\sin(k_0 |\rr_1-\rr_2|)}{k_0|\rr_1-\rr_2|}\ldots \frac{\sin(k_0 |\rr_m-\rr_1|)}{k_0|\rr_m-\rr_1|},
\end{split}
\ee
where $S_L$ is the volume of a sphere of radius $L$.
It is convenient to use the following expansion of the integral kernel~\citep[Eq. 8.533]{S_Gradshteyn7th}
\be\label{eq:3DKernel_Expansion}
	\frac{\sin(k_0|\rr-\rr'|)}{4\pi k_0|\rr-\rr'|} =  \sum_{l=0}^\infty \sum_{m=-l}^lj_l(k_0r) Y_{lm}(\nn)Y_{lm}^*(\nn') j_l(k_0r'),
\ee
where $r\equiv |\rr|$ and $\nn = \rr/r = (\sin\theta\cos\phi,\sin\theta\sin\phi,\cos\theta)^T$.
Substituting \eqnref{eq:3DKernel_Expansion} into \eqnref{eq:Lambda_3D} and using the orthogonality condition $\int \text{d}\nn \,Y_{lm}^*(\nn)Y_{kn}(\nn)=\delta_{lk}\delta_{mn}$, we have
\be\label{eq:Lambda_3D_expansion}
    T_m^\text{3D} = (4\pi \rho_\text{3D})^m  \sum_{l=0}^\infty  (2l+1)\cpare{\int_0^L\!\!\text{d}r\, r^2 [j_l(k_0r)]^2}^m \equiv (4\pi \rho_\text{3D})^m \sum_{l=0}^\infty(2l+1)\gamma_l^m,
\ee
where we defined
\be\label{eq:gamma_l_3D}
\gamma_l \equiv \frac{L^{3}}{2}\spare{j_l^2(k_0L)-j_{l-1}(k_0L)j_{l+1}(k_0L)}.
\ee
Depending on the value of $l$, \eqnref{eq:gamma_l_3D} has different asymptotic behavior for a large argument $k_0L$. For $l\ll \sqrt{k_0L}$, we can approximate $\gamma_l \simeq 3N/8\rho_\text{3D}\pi(k_0L)^2$. Instead, when $l\gg (k_0L)^2$, the function decays superexponentially with $l$ and, to leading order, we approximate $\gamma_l\simeq 0$. 
We can thus write \eqnref{eq:Lambda_3D_expansion} as
\be\label{eq:T_m_3D_step}
\begin{split}
    T_m^\text{3D} = \pare{\frac{3 N}{2(k_0L)^2}}^m \sum_{l=0}^{\lfloor \sqrt{k_0L} \rfloor} (2l+1) + \sum_{l=\lfloor \sqrt{k_0L} \rfloor}^{\lfloor (k_0L)^2 \rfloor}(2l+1)\gamma_l^m.
\end{split}
\ee
At this point we note that \eqnref{eq:gamma_l_3D} decreases from a maximum value at $l \simeq 0$ to zero when $l\approx \lfloor k_0L \rfloor$ [\figref{fig:Plot_Gamma_l_3D}]. 
The remaining terms of the sum for $l = \lfloor \sqrt{k_0L} \rfloor, \ldots , \lfloor (k_0L)^2 \rfloor$, which are very small, can be upper-bounded by $\gamma_l \simeq 3N/8\rho_\text{3D}\pi(k_0L)^2$, leading to
\be\label{eq:T_m_3D_step}
\begin{split}
    T_m^\text{3D} \leq \pare{\frac{3 N}{2(k_0L)^2}}^m(\lfloor (k_0L)^2 \rfloor +1)^2.
\end{split}
\ee
Finally, we can upper-bound \eqnref{eq:Lambda_3D_expansion} as
\be
    T_m^\text{3D} \leq \pare{\frac{3 N}{2(k_0L)^2}}^m(\lfloor (k_0L)^2 \rfloor +1)^2 \simeq \pare{\frac{3 }{2(k_0d)^2}}^m(\lfloor (k_0L)^2 \rfloor +1)^2 N^{\frac{m}{3}},
\ee
where we used $L=N^{1/3}d$.
Substituting this result into \eqnref{eq:Gamma_max_limit_trace} we obtain
\be
    \frac{\Gamma_\text{max}}{\Gamma_0}  \leq \frac{3 N^{1/3}}{2(k_0 d)^2}\sim N^{\frac{1}{3}}.
\ee

\section{Eigenvalues and Eigenvectors of $\boldsymbol{\Gamma}$}

In this section, we show that the functions in \eqnref{eq:Ansatz_Eigenstates} in the main text are a set of eigenmodes of \eqnref{eqS:Fredholm_Eq}.
For the 3D case, these constitute the full set of eigenmodes and eigenvalues. For 1D and 2D, they are only a subset. We show, however, that this subset contains eigenvectors whose associated eigenvalue (decay rate) has the same scaling as $\Gamma_\text{max}$, derived in \secref{sec:Scaling_Gamma_max_Clouds} above.

\subsection{1D Clouds}

We proceed to solve the eigenvalue problem for a 1D cloud in the continuum limit to obtain an eigenvector whose associated eigenvalue scales like $\Gamma_\text{max}$. For a 1D cloud, equation~\eqref{eqS:Fredholm_Eq} reads 
\be\label{eq:Eval_Problem_1D}
    \int_{-L/2}^{L/2}\!\! \text{d}y\,\rho_\text{1D}\frac{\sin(k_0|x-y|)}{k_0 |x-y|} \psi(y) = \frac{\Gamma}{\Gamma_0} \psi(x),
\ee
where the even symmetry of the kernel allows us to write $|x-y|\rightarrow x-y$.
To solve \eqnref{eq:Eval_Problem_1D}, we use the expansion of a spherical wave (see \citep[Eq.8.533 on pp. 940]{S_Gradshteyn7th}) to write the kernel as
\be\label{eq:Kernel_Expansion_1D}
    \frac{\sin(k_0(x-y))}{k_0 (x-y)} = \sum_{n=0}^\infty (2n+1) j_n(k_0x)j_n(k_0 y),
\ee
where $j_n(k_0x)$ is the spherical Bessel function of the first kind of order $n$.
Equation~(\ref{eq:Kernel_Expansion_1D}) suggests to look for eigenvectors of the form [see~\eqnref{eq:Ansatz_Eigenstates}]
\be\label{eq:EigVec_Ansatz_1D}
    \psi_l(x) = \mathcal{N}_lj_l(k_0 x),
\ee
where $\mathcal{N}_l$ is a dimensionless normalization constant and $l$ is an integer number. Substituting Eq.~(\ref{eq:EigVec_Ansatz_1D}) into Eq.~(\ref{eq:Eval_Problem_1D}) and using the expansion Eq.~(\ref{eq:Kernel_Expansion_1D}), we obtain
\be\label{eq:Sum_1D_Integral}
    \sum_{n=0}^\infty (2n+1)\rho_\text{1D}\psi_n(x) \underbrace{\int_{-L/2}^{L/2}\!\! \text{d}y\, j_n(k_0y) j_l(k_0y)}_{\equiv\,\mathcal{I}_\text{1D}} = \frac{\Gamma_l}{\Gamma_0} \psi_l(x),
\ee
where we swapped the order of integral and sum. 
To evaluate the integral $\mathcal{I}_\text{1D}$, we use the following identity
\be\label{eq:Identity_Spherical_Bessel}
    j_n(x) = \frac{\im^{-n}}{2}\int_{-1}^1 \text{d}u\, e^{\im ux} P_n(u),    
\ee
where $P_n(u)$ is the Legendre polynomial of order $n$. Substituting \eqnref{eq:Identity_Spherical_Bessel} into $\mathcal{I}_\text{1D}$ and performing the integral over $y$, we obtain
\be\label{eq:Integral_two_spherical_bessel_j}
\begin{split}
    \mathcal{I}_\text{1D} = \frac{\im^{-n-l}}{2 k_0} \int_{-1}^1 \text{d}u\text{d}v\, \frac{\sin[(u+v)\frac{k_0L}{2}]}{u+v}P_n(u)P_l(v)
    = \frac{\im^{-n}(-\im)^{-l}}{2 k_0} \int_{-1}^1 \text{d}u\text{d}v\, \frac{\sin[(u-v)\frac{k_0L}{2}]}{u-v}P_n(u)P_l(v),
\end{split}
\ee
where in the last step we used $P_l(-v)=(-1)^lP_l(v)$. 
The function $\sin[(u-v)k_0L/2]/(u-v)$ is sharply peaked around $u=v$ with a characteristic width of $\approx (k_0L)^{-1}$. Consequently, we can employ the approximation $\sin[(u-v)k_0L/2]/(u-v) \simeq \pi \delta(u-v)$ whenever $(k_0L)^{-1}$ is much smaller than the typical scale over which $P_n(u)$ varies. For large $n$, Legendre polynomials exhibit rapid variation near the boundaries at $u=\pm 1$. As demonstrated by Tricomi~\cite{S_Tricomi1950}, the distance between $u=\pm 1$ and the nearest zeros of $P_n(u)$ scales as $n^{-2}$. Therefore, this delta function approximation is valid provided that $k_0L \gg n^2$.
Substituting this result back into \eqnref{eq:Sum_1D_Integral} we have
\be
    \int_{-L/2}^{L/2}\!\! \text{d}y\,\rho_\text{1D} \frac{\sin(k_0(x-y))}{k_0 (x-y)} \psi_l(y) = \frac{\pi}{k_0d}\, \psi_l(x).
\ee
This proves that in the limit of large arrays ($k_0L\gg1$), a set of eigenvalues and eigenvectors of the system is given by
\be
    \psi_l(x) = \mathcal{N}_lj_l(k_0x),\qquad \Gamma_l =  \Gamma_0\frac{\pi\rho_\text{1D}}{k_0}.
\ee
We remark that the functions $\{\psi_l(x)\}_l$ are orthogonal only in the limit of infinite system size $k_0L\rightarrow \infty$.

To conclude the analysis for 1D clouds, we compute the normalization constant $\mathcal{N}_l$. We require the eigenmode $\psi_n(x)$ to be normalized, such that
\be
    \int_{-L/2}^{L/2}\!\!\text{d}y\, \rho_\text{1D}|\psi_l(y)|^2 = \mathcal{N}_l^2 \rho_\text{1D} \int_{-L/2}^{L/2}\!\!\text{d}y\, |j_l(k_0y)|^2 = 1.
\ee
The integral of the spherical Bessel function can be evaluated as in $\mathcal{I}_\text{1D}$. As before, we approximate $\sin(x k_0L/2)/x\approx \pi\delta(x)$ in the limit $k_0L\gg l^2$ and finally obtain
\be
    \mathcal{N}_l = \sqrt{\frac{2l+1}{\pi}\frac{k_0}{\rho_\text{1D}}}.
\ee

\subsection{2D Clouds}

We now proceed to solve the eigenvalue problem for a 2D cloud of atoms. For this case, Equation~\eqref{eqS:Fredholm_Eq} reads 
\be\label{eq:Eval_Problem_2D}
    \mathcal{I}_\text{2D} \equiv \int\!\! \text{d}^2\rr'\, \rho_\text{2D} \frac{\sin(k_0|\rr-\rr'|)}{k_0 |\rr-\rr'|} \psi(\rr') = \frac{\Gamma}{\Gamma_0}\psi(\rr).
\ee
To evaluate \eqnref{eq:Eval_Problem_2D} we use the following representation of the kernel~\citep[Eq.~(15.8)]{S_Janaswamy2020}
\be\label{eq:Kernel2D_Expansion}
\begin{split}
    \frac{e^{-\im k_0|\rr-\rr'|}}{ |\rr-\rr'|} = \frac{1}{\im}\int_0^\infty\!\!\! \frac{\text{d}k\, k}{\sqrt{k_0^2-k^2}}\!\! \sum_{n=-\infty}^\infty\!\!
    J_n(k r)J_n(k r')e^{\im n(\phi-\phi')}.
\end{split}
\ee
In order to use the expansion in \eqnref{eq:Kernel2D_Expansion} it is convenient to write \eqnref{eq:Eval_Problem_2D} as
\be\label{eq:Integral_Split_Kernet}
\begin{split}
    \mathcal{I}_\text{2D} = \int\!\!\text{d}^2\rr'\, \rho_\text{2D}\frac{e^{-\im k_0|\rr-\rr'|}}{-2\im k_0 |\rr-\rr'|}\psi(\rr')
    + \spare{\int\!\!\text{d}^2\rr'\,\rho_\text{2D} \frac{e^{-\im k_0|\rr-\rr'|}}{-2\im k_0 |\rr-\rr'|}\psi^*(\rr')}^* \equiv I_A + I_B.
\end{split}
\ee
The form of the kernel suggests to look for eigenstates of the form
\be\label{eq:EigVec_Ansatz_2D}
    \psi_m(\rr) \equiv \mathcal{N}_m J_m(k_0r) e^{\im m\phi},
\ee
where $J_m(x)$ is the Bessel function of the first kind of integer order $m$. Note that \eqnref{eq:EigVec_Ansatz_2D} form a set of orthogonal states labeled by the integer $m$.
Since the two integrals $I_A$ and $I_B$ are very similar we discuss in detail how to compute the former only. The derivation for the latter proceeds analogously. 
Substituting the expansion in \eqnref{eq:Kernel2D_Expansion} into the first integral in \eqnref{eq:Integral_Split_Kernet} we have
\be\label{eq:I_A_angular_integrated}
\begin{split}
    I_\text{A} =&  \mathcal{N}_m\frac{\rho_\text{2D}}{2k_0}\int_0^L\!\! \text{d}r'\,r'\int_0^{2\pi}\!\!\text{d}\phi'\int_0^\infty\!\! \frac{\text{d}k\, k}{\sqrt{k_0^2-k^2}} \sum_{n=-\infty}^\infty J_n(k r)J_n(k r')e^{\im n\phi} e^{\im (m-n)\phi'}J_m(k_0r')\\
    = &  \mathcal{N}_m \frac{\pi\rho_\text{2D}}{k_0}e^{\im m\phi} \int_0^\infty\!\! \frac{\text{d}k\, k}{\sqrt{k_0^2-k^2}} J_m(k r)\int_0^L\!\! \text{d}r'\,r' J_m(k r')J_m(k_0r'),
\end{split}
\ee
where in the last step we used  $\int_0^{2\pi} \exp[\im (m-n)\phi'] = 2\pi \delta_{nm}$ and  swapped the order of the radial and momentum-space integrals.
The radial integral is tabulated in Ref.~\citep[Eq.6.521.1 on pp. 664]{S_Gradshteyn7th} and reads
\be\label{eq:Radial_Integral}
    f_m(k,k_0,L)\equiv\int_0^L\!\! \text{d}r'\,r'J_m(k_0 r')J_m(k r') = \frac{k_0L J_{m-1}(k_0L)J_m(kL)-kL J_{m-1}(kL)J_m(k_0L)}{k^2-k_0^2}.
\ee
Substituting this result back into \eqnref{eq:I_A_angular_integrated}, we obtain
\be\label{eq:I_A_final}
\begin{split}
    I_\text{A} =& \mathcal{N}_m \frac{\pi\rho_\text{2D}}{k_0}e^{\im m\phi}\! \int_0^\infty\!\! \frac{\text{d}k\, k}{\sqrt{k_0^2-k^2}} J_m(k r)f_m(k,k_0,L) \simeq  \pare{\frac{\pi\rho_\text{2D}}{k_0} \int_0^\infty\!\! \frac{\text{d}k\, k}{\sqrt{k_0^2-k^2}} f_m(k,k_0,L)}\psi_m(r)\equiv \Lambda_m(L) \psi_m(r),
\end{split}
\ee
where in the second step we approximated $J_m(kr)\simeq J_m(k_0r)$ using the fact that $k/\sqrt{k_0^2-k^2}$ is sharply peaked around $k=k_0$ and, in the limit of large system size $k_0L\gg 1$, one has $\lim_{k_0L\rightarrow \infty} f_m(k,k_0,L) = \delta(k-k_0)/k$. Below, we discuss the accuracy of the approximation in \eqnref{eq:I_A_final}.
The calculation of the second integral in \eqnref{eq:Integral_Split_Kernet} proceeds similarly and leads to the result
\be\label{eq:I_B_final}
    I_\text{B} \simeq  \spare{\pare{\frac{\pi \rho_\text{2D}}{k_0} \int_0^\infty\!\! \frac{\text{d}k\, k}{\sqrt{k_0^2-k^2}} f_m(k,k_0,L)}\psi^*_m(r)}^*\equiv \Lambda_m(L)^* \psi_m(r).
\ee

Combining \eqnref{eq:I_A_final} and \eqnref{eq:I_B_final} into \eqnref{eq:Eval_Problem_2D}, we obtain that in the limit of large arrays ($k_0L\gg1$) the functions in \eqnref{eq:EigVec_Ansatz_2D} are eigenvectors of the system with associated eigenvalues 
\be\label{eq:Gamma_m_2D}
	\Gamma_m = 2\Re[\Lambda_m(L)] = \frac{2\pi}{k_0^2}\rho_\text{2D} \Re\spare{\int_0^\infty\!\! \text{d}k\,\frac{ k f_m(k,k_0,L)}{\sqrt{1-(k/k_0)^2}}}.
\ee

\begin{figure}
	\includegraphics[width=0.8\columnwidth]{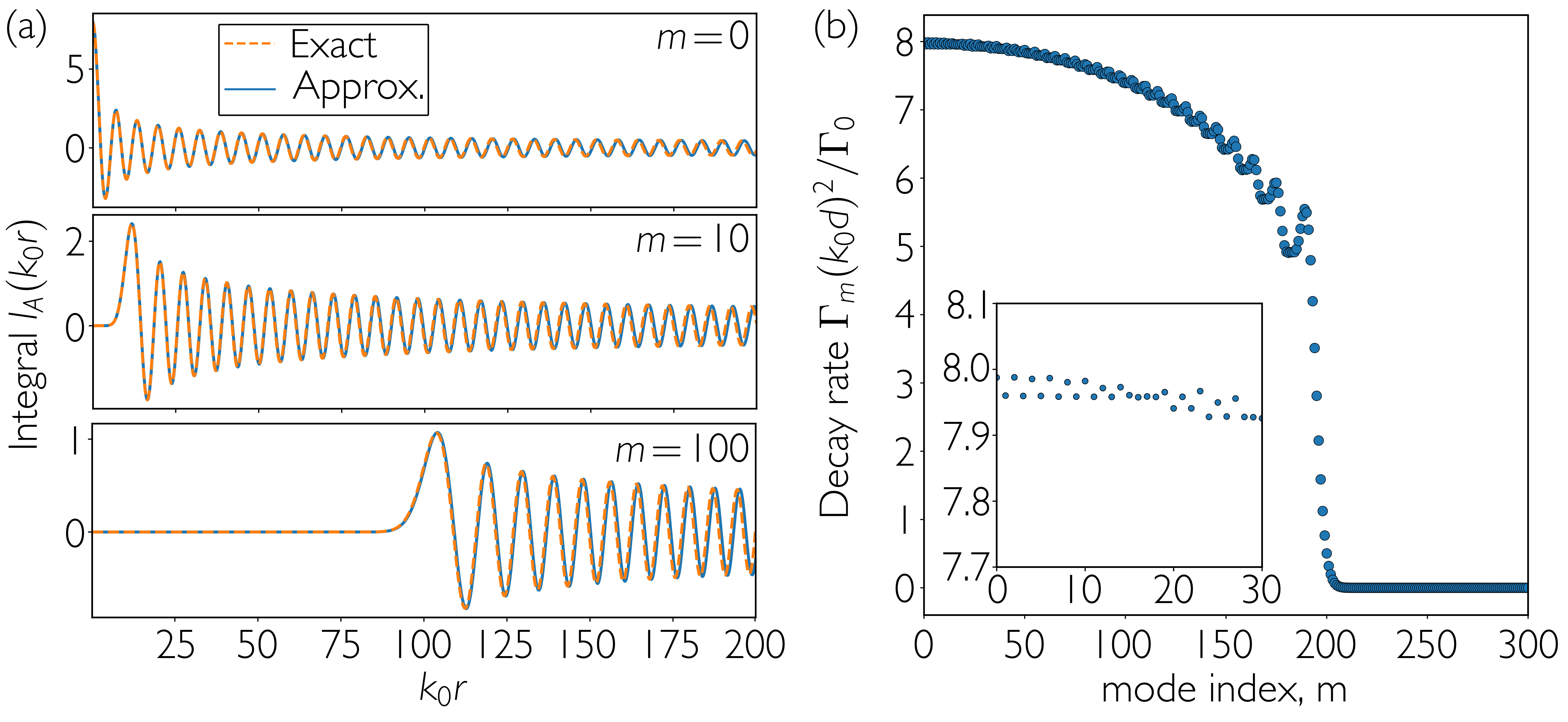}
	\caption{(a) Comparison between $I_\text{A}^{\text{exact}}(k_0r)$ (dashed orange line) and $I_\text{A}^{\text{approx}}(k_0r)$ (solid blue line) for three different values of $m$ as indicated in each panel for a 2D cloud with $k_0 L=200$. (b) Plot of the eigenvalues $\Gamma_m/\Gamma_0$ as defined in \eqnref{eq:Gamma_2D_closed_form} as a function of the mode index $m$. Inset: particular of the dependence at small mode index.}\label{fig:FigS3}
\end{figure}

Let us now prove that in the limit of large system size $k_0L\gg 1$, the decay rate in \eqnref{eq:Gamma_m_2D} scales as $\sim \sqrt{k_0L}$.
We need to compute the real part of the following integral
\be\label{eq:Step_Lambda_nL}
    \mathcal{J}_m \equiv \int_0^\infty \frac{\text{d}x\, x}{\sqrt{1-x^2}} \frac{\ell J_{m-1}(\ell)J_m(x\ell)-x\ell J_{m-1}(x\ell)J_m(\ell)}{x^2-1},
\ee
which is obtained from the integral in \eqnref{eq:Gamma_m_2D} by the change of variables $x\equiv k/k_0$ and by defining $\ell \equiv k_0 L$.
Note that the pole at $x=1$ is integrable as for $x\sim 1$ the integrand can be approximated as
\be
    -x\frac{\ell J_{m-1}(\ell)J_m(x\ell)-x\ell J_{m-1}(x\ell)J_m(\ell)}{(1-x^2)^{3/2}}  \simeq - \frac{\ell}{2\sqrt{2}} \frac{J_m^2(\ell)-J_{m-1}^2(\ell)J_{m+1}^2(\ell)}{\sqrt{1-x}} + O(1-x).
\ee
To evaluate this integral, it is convenient to rewrite it as
\be\label{eq:Lambda_L_integrals}
\begin{split}
    \mathcal{J}_m  =& -\int_0^1\!\! \text{d}x \frac{\text{d}}{\text{d}x}\pare{\inv{\sqrt{1-x^2}}} \spare{\ell J_{m-1}(\ell)J_m(x\ell)-x\ell J_{m-1}(x\ell)J_m(\ell)}\\
    &+ \im\int_1^\infty\!\! \text{d}x \frac{\text{d}}{\text{d}x}\pare{\inv{\sqrt{x^2-1}}} \spare{\ell J_{m-1}(\ell)J_m(x\ell)-x\ell J_{m-1}(x\ell)J_m(\ell)}.
\end{split}
\ee
We proceed to evaluate the integral over the unit interval $[0,1]$.
Integrating by parts, we obtain
\be\label{eq:U_01}
    U_{[0,1]} \equiv \ell J_{m-1}(\ell)J_m(0) + \spare{\ell J_{m-1}(\ell) \pare{\int_0^1\!\!\text{d}x\,\frac{ 1}{\sqrt{1-x^2}}\frac{\text{d}}{\text{d}x}J_m(x\ell)} - \ell J_{m}(\ell)\pare{\int_0^1\!\!\text{d}x\, \frac{1}{\sqrt{1-x^2}}\frac{\text{d}}{\text{d}x}[xJ_{m-1}(x\ell)]}},
\ee
where the boundary term vanishes in the limit $x\rightarrow 1$.
Using the relation for the derivative of the Bessel functions $\text{d}J_m(x \ell)/\text{d}x= \ell[J_{m-1}(x\ell)-J_{m+1}(x\ell)]/2$, these integrals can be reduced to known integrals and read (see Ref.~\citep[Eq.~(6.552.4) on pp. 674]{S_Gradshteyn7th}),
\bea
    \int_0^1\!\!\frac{\text{d}x}{\sqrt{1-x^2}}\frac{\text{d}}{\text{d}x}J_m(x\ell) &=& \frac{\pi \ell}{4}\spare{J^2_{\frac{m-1}{2}}\pare{\frac{\ell}{2}}-J^2_{\frac{m+1}{2}}\pare{\frac{\ell}{2}}},\\
    \int_0^1\!\! \frac{\text{d}x}{\sqrt{1-x^2}}\frac{\text{d}}{\text{d}x}[xJ_{m-1}(x\ell)] &=& \frac{\pi}{2} J_{\frac{m-1}{2}}\pare{\frac{\ell}{2}} + \frac{\pi\ell}{2}\frac{\text{d}}{\text{d}\ell} \spare{J^2_{\frac{m-1}{2}}\pare{\frac{\ell}{2}}}.
\eea
Substituting back these results into \eqnref{eq:U_01}, we obtain
\be
    U_{[0,1]} = \frac{\pi\ell^2}{4}J_{m-1}(\ell)\spare{J^2_{\frac{m-1}{2}}\pare{\frac{\ell}{2}}-J^2_{\frac{m+1}{2}}\pare{\frac{\ell}{2}}} - \frac{\pi \ell}{2}J_m(\ell)\cpare{J^2_{\frac{m-1}{2}}\pare{\frac{\ell}{2}} + \ell \frac{\text{d}}{\text{d}\ell}\spare{J^2_{\frac{m-1}{2}}\pare{\frac{\ell}{2}}}}.
\ee
Following a similar procedure, we prove that the second integral in \eqnref{eq:Lambda_L_integrals} gives a purely imaginary contribution and hence it does not contribute to \eqnref{eq:Gamma_m_2D}.
We can thus write the decay rate in \eqnref{eq:Gamma_m_2D} as
\be\label{eq:Gamma_2D_closed_form}
	\Gamma_m = \frac{2\pi}{k_0^2}\rho_\text{2D}\cpare{\frac{\pi\ell^2}{4}J_{m-1}(\ell)\spare{J^2_{\frac{m-1}{2}}\pare{\frac{\ell}{2}}-J^2_{\frac{m+1}{2}}\pare{\frac{\ell}{2}}} - \frac{\pi \ell}{2}J_m(\ell)\pare{J^2_{\frac{m-1}{2}}\pare{\frac{\ell}{2}} + \ell \frac{\text{d}}{\text{d}\ell}\spare{J^2_{\frac{m-1}{2}}\pare{\frac{\ell}{2}}}}}.
\ee
In the large system limit, for $m^2\ll k_0L$, we can approximate \eqnref{eq:Gamma_m_2D} to leading order in $k_0L$ as
\be\label{eq:Scaling_Gamma_m_2D}
	\Gamma_m \simeq \frac{\lambda^2_0\rho_\text{2D}}{2\pi}\sqrt{\frac{k_0L}{\pi}}\sim N^{1/4},
\ee
where in the last passage we used that at fixed density $k_0L = \sqrt{ Nk_0^2/\pi\rho_\text{2D}}\sim N^{1/2}$.
In this limit $\Gamma_m$ is independent on $m$ as confirmed numerically in \figref{fig:FigS3}(b), which also shows that $\Gamma_m=0$ for $m\gtrsim k_0L$.

This results rests on the validity of the approximation $J_m(kr)\simeq J_m(k_0r)$ in \eqnref{eq:I_A_final}. We now verify its accuracy by comparing the following functions
\bea
    I_\text{A}^{\text{exact}}(k_0r) &\equiv& \int_0^1\!\!\text{d}x \frac{x}{\sqrt{1-x^2}} \tilde{f}_m(x\ell,\ell) J_m(x k_0 r),\label{eq:I_A_Exact}\\
    I_\text{A}^{\text{approx}}(k_0r) &\equiv& \pare{\int_0^1\!\!\text{d}x \frac{x}{\sqrt{1-x^2}} \tilde{f}_m(x\ell,\ell)} J_m(k_0 r),\label{eq:I_A_Approximated}
\eea
where $\tilde{f}_m(x\ell,\ell)\equiv [\ell J_{m-1}(\ell)J_m(x\ell)-x\ell J_{m-1}(x\ell)J_m(\ell)]/(x^2-1)$ is obtained from $f_m(k,k_0,L)$ in \eqnref{eq:Radial_Integral} after the change of variable $x=k/k_0$ and $\ell=k_0L$. Here, we only consider the integration to run over the interval $x\in[0,1]$, as this is the only part that contributes to $\Gamma_m$.
We confirm the validity of the approximation from the very good agreement between \eqnref{eq:I_A_Exact} and \eqnref{eq:I_A_Approximated} [see~\figref{fig:FigS3}(a)].

To conclude the analysis, we compute the normalization constant in \eqnref{eq:EigVec_Ansatz_2D}. The eigenmodes of the system are normalized such that 
\be
\begin{split}
    \rho_\text{2D}\int \!\!\text{d}\rr \, |\psi_m(\rr)|^2 =& \mathcal{N}_m^2 2\pi \rho_\text{2D}\int_0^L \text{d}r\,r J_m^2(k_0r)= \frac{\pi\rho_\text{2D}L}{k_0}\mathcal{N}_m^2 \cpare{k_0L \spare{J_m^2(k_0L)+J_{m+1}^2(k_0L)}-2m J_m(k_0L)J_{m+1}(k_0L)}\\
    =& 1.
\end{split}
\ee
Taking the leading order term in the expansion for $k_0L\gg m^2$, we obtain
\be
    \mathcal{N}_m \simeq \sqrt{\frac{k_0}{2L\rho_\text{2D}}}.
\ee

\subsection{3D Clouds}

The solution of the eigenvalue problem in \eqnref{eqS:Fredholm_Eq} has been obtained before for uniformly distributed atoms in a sphere~\cite{S_Svidzinsky2008}. We repeat here the argument for completeness. For a 3D cloud of atoms, \eqnref{eqS:Fredholm_Eq} takes the form
\be\label{eq:Eval_Problem_3D}
    \int\!\! \text{d}^3\rr'\,\rho_\text{3D} \frac{\sin(k_0|\rr-\rr'|)}{k_0 |\rr-\rr'|} \psi(\rr') = \frac{\Gamma}{\Gamma_0}\psi(\rr).
\ee
For the eigenfunctions of the system we choose the following ansatz
\be\label{eq:ansatz_3D}
	\psi_{l,m}(\rr)=\mathcal{N}_l j_l(k_0 r) Y_{lm}(\nn),
\ee
where $\rr=\nn r$, and $\nn$ is a unit vector.
Substituting \eqnref{eq:ansatz_3D} into \eqnref{eq:Eval_Problem_3D}, using the kernel expansion in \eqnref{eq:3DKernel_Expansion}, and the orthogonality of spherical harmonics, it is straightforward to show that \eqnref{eq:ansatz_3D} are eigenfunctions of \eqnref{eq:Eval_Problem_3D} with eigenvalue 
\be
	\frac{\Gamma_{lm}}{\Gamma_0} = 4\pi \rho_\text{3D}\int_0^L\!\!\!\text{d}r\, r^2 [j_l(k_0 r)]^2 = 2\pi\rho_\text{3D}L^3 \spare{j_l^2(k_0L) - j_{l-1}(k_0L)j_{l+1}(k_0L)} \simeq 2\pi \frac{\rho_\text{3D}}{k_0^3} k_0L \sim N^{\frac{1}{3}},
\ee
where we took the leading order in the asymptotic expansion for $k_0L\gg l^2$.

We conclude this section by computing the normalization constant in \eqnref{eq:ansatz_3D}. Imposing that the eigenmodes are normalized we have
\be
    \mathcal{N}_{l} = \cpare{\rho_\text{3D}L^3 \spare{j_l^2(k_0L)-j_{l-1}(k_0L)j_{l+1}(k_0L)}}^{-\frac{1}{2}} \simeq \sqrt{\frac{k_0^2}{\rho_\text{3D}L}},
\ee
where in the last step we assumed once again $k_0L\gg l^2$.

\section{Distribution of emitted photons from an atomic ensemble}

In this section we compute the emission pattern $\mu(\phi,\theta)$ for the photons emitted from an ordered or disordered 1D, 2D, and 3D ensemble of atoms in free space. The general form of the emission pattern from an ensemble of atoms in a state $\ket{\psi}$ reads
\be\label{eq:Emission_Pattern_General}
    \mu(\phi,\theta) \equiv \Big\vert \inv{N} \int_{\mathbb{R}^3}\!\!\! \text{d} \rr\, \rho_D(\rr) e^{-\im k_0 \uu\cdot \rr} \psi(\rr)\Big\vert^2,
\ee
where $\psi(\rr)=\braket{\rr}{\psi}$ is the atomic wavefunction in position space, $N = (N_\text{1D})^D$ is the total number of atoms in the ensemble, $\uu \equiv (\cos\phi\sin\theta,\sin\phi\sin\theta,\cos\theta)$, and we introduced the atom density $\rho_D(\rr)$. For atomic arrays, the density reads
\be\label{eq:Density_Array}
	\rho_D^\text{array}(\rr) = \sum_{j=1}^N
	\left\{
	\begin{array}{ll}
	\delta(x)\delta(y) \delta(z-d j_z) & \quad \text{for $D=1$},\\
	\delta(x-dj_x)\delta(y-dj_y) \delta(z) & \quad \text{for $D=2$},\\
    \delta(x-dj_x) \delta(y-dj_y) \delta(z-dj_z) &\quad \text{for $D=3$},
	\end{array}
	\right .
\ee
while for a disordered atomic ensemble uniformly distributed in a sphere ($D=3$), disk ($D=2$), or line ($D=1$) the density reads
\be\label{eq:Density_Cloud}
    \rho_D^\text{cloud}(\rr) = \rho_D
	\left\{
	\begin{array}{ll}
	\delta(x)\delta(y) & \quad \text{for $D=1$ and $-L/2\leq z\leq L/2$},\\
	\delta(z) & \quad \text{for $D=2$ and $\sqrt{x^2+y^2}\leq L$},\\
    1 &\quad \text{for $D=3$ and $\sqrt{x^2+y^2+z^2}\leq L$},
	\end{array}
	\right .
\ee
where $\rho_D\equiv\frac{N}{V}$ and $V$ is the volume of the $D$-dimensional cloud.

The angular distribution of the emitted photon is also determined by the atomic polarization, $\mathcal{D}(\uu)$. In the following we are interested in the scaling with the size of the system of the total solid angle $\Delta\W$ into which photons are emitted. The dipole emission pattern does not change such scaling, thus we consider the scalar case of $\mathcal{D}(\W)=1/4\pi$.

\subsection{Ordered Arrays}

The principal eigenvectors in ordered arrays are well approximated by the spin-wave states~\cite{S_Clemens2003,S_Shahmoon2017,S_Perczel2017, S_Zhang2019,S_AsenjoGarcia2017PRX},
\be\label{eq:principal_evect_array}
    \ket{\kk_0} \equiv \frac{1}{\sqrt{N}}\sum_{\jj} e^{\im \kk_0 \cdot \rr_\jj} \spl_{\jj} \ket{g\ldots g} 
\ee
where $\rr_\jj=d \jj$ is the lattice position of the $j$-th atom. We note that there exist several directions of $\kk_0$ for which the states in \eqnref{eq:principal_evect_array} are degenerate. It is thus convenient to fix a direction. 
Substituting \eqnref{eq:principal_evect_array} into \eqnref{eq:Emission_Pattern_General}, using \eqnref{eq:Density_Array} and assuming a square lattice, we reduce the integral to a sum of complex exponentials evaluated at different lattice points,
\be\label{eq:mu_array_general}
\begin{split}
    \mu^\text{array}(\phi,\theta) =& \bigg\vert \frac{1}{N}\sum_{j_x,j_y,j_z=1}^{N_\text{1D}}\exp\cpare{\im k_0d[j_x(k_{0x}-u_x) +j_y(k_{0y}-u_y) +j_z(k_{0z}-u_z)]}\bigg\vert^2\\
    =& \bigg\vert \frac{1}{N_\text{1D}}\sum_{j_x=1}^{N_\text{1D}}e^{\im k_0 j_x (k_{0x}-u_x)}\bigg\vert^2\bigg\vert \frac{1}{N_\text{1D}}\sum_{j_y=1}^{N_\text{1D}}e^{\im k_0 j_y (k_{0y}-u_y)}\bigg\vert^2\bigg\vert \frac{1}{N_\text{1D}}\sum_{j_z=1}^{N_\text{1D}}e^{\im k_0 j_z (k_{0z}-u_z)}\bigg\vert^2\\
    \equiv& \mu_x(\phi,\theta)\mu_y(\phi,\theta)\mu_z(\phi,\theta)
\end{split}
\ee
We now proceed to analyze the emission pattern in \eqnref{eq:mu_array_general} for arrays of different dimensionality.

{\bf 1D Arrays.} For one dimensional arrays along the $\uez$ axis, we have $\kk_0 = k_0\uez$. The emission pattern then reads
\be\label{eq:mu_z}
    \mu_\text{1D}^\text{array}(\phi,\theta) = \mu_z(\phi,\theta) = \inv{N^2}\pare{\frac{\sin[k_0L(1-\cos\theta)/2]}{\sin[k_0d(1-\cos\theta)/2]}}^2,
\ee
where $N=N_\text{1D}$. The emission from the state given by \eqnref{eq:principal_evect_array} with $\kk_0=k_0 \uez$ is symmetric around the array's axis and concentrated within a small angle $\theta$ around $\uez$. 
The emission pattern is concentrated within a cone with angle $\Delta\theta$ that can be estimated as the separation between the first two zeros of \eqnref{eq:mu_z} as
\be\label{eq:Delta_Theta}
    \Delta \theta = 2\sqrt{\frac{4\pi}{k_0L}}\sim N^{-\frac{1}{2}}.
\ee
The solid angle into which a photon is emitted can thus be estimated as
\be
    \Delta\W_{\text{1D}} \simeq (\Delta\theta)^2 = \frac{16\pi}{k_0L} \sim N^{-1}.
\ee
As a consequence, the scaling of the maximum emission rate for a 1D array is given by $R_\star/\Gamma_0\sim N^2\times \Delta \W_\text{1D}\sim N$.

{\bf 2D Arrays.} We consider a two dimensional square-lattice in the $xy$-plane where the array's axes are aligned along the  $\uex$ and $\uey$ axis. We also assume $\kk_0 = k_0\uex$ in \eqnref{eq:principal_evect_array}. The emission pattern reads
\be\label{eq:mu_2D_array}
    \mu_{\text{2D}}^{\text{array}}(\theta,\phi) = \mu_x(\theta,\phi)\mu_y(\theta,\phi) = \inv{N}\pare{\frac{\sin[k_0L (1-\cos\phi\sin\theta)/2]}{\sin[k_0 d(1-\cos\phi\sin\theta)/2]}}^2 \inv{N}\pare{\frac{\sin(\frac{k_0L}{2}\sin\phi\sin\theta)}{\sin(\frac{k_0 d}{2}\sin\phi\sin\theta)}}^2,
\ee
where $N=N_\text{1D}^2$. Photon emission is collimated around the $\uex$ axis.
Expanding the argument of the sine functions in \eqnref{eq:mu_2D_array} around $\phi=0$ and $\theta=\pi/2$ up to second order yields
\be\label{eq:mu_2D_array_small_angle}
    \mu_{\text{2D}}^{\text{array}}(\theta,\phi) \simeq \inv{N}\pare{\frac{\sin[k_0L (\phi^2+\tilde{\theta}^2)/4]}{\sin[k_0 d(\phi^2+\tilde{\theta}^2)/4]}}^2\inv{N}\pare{\frac{\sin(k_0L\phi/2)}{\sin(k_0 d\phi/2)}}^2,
\ee
where $\tilde{\theta}\equiv \theta-\pi/2$. We set $\tilde{\theta}=0$ in \eqnref{eq:mu_2D_array_small_angle}, and compute the angle $\Delta\phi$ (into which photons are predominantly emitted in the $xy$-plane) as the distance between the first two zeros of $\mu_{\text{2D}}^{\text{array}}(\phi,\pi/2)$. The two terms in \eqnref{eq:mu_2D_array_small_angle} yield two separate conditions. The most restrictive is given by the last term in \eqnref{eq:mu_2D_array_small_angle} and implies
\be
    \Delta\phi = \frac{4\pi}{k_0L} \simeq N^{-1}.
\ee
Similarly, photons are mostly emitted within a small angle $\Delta\theta$ in the $xz$-plane. This angle is computed as the distance between the first two zeros of $\mu_{\text{2D}}^{\text{array}}(0,\theta)$ 
and is given in \eqnref{eq:Delta_Theta}.
The solid angle into which a photon is emitted by a 2D array can thus be estimated as
\be
    \Delta\W_{\text{2D}} \simeq \Delta\phi\Delta\theta = 2\pare{\frac{4\pi}{k_0L}}^{\frac{3}{2}} \sim N_\text{1D}^{-\frac{3}{2}}.
\ee
This scaling of the solid angle with the atom number confirms the recent numerical results obtained in Ref.~\cite{S_vonMilczewski2025}.
As a consequence, the scaling of the maximum emission rate for a 2D array is given by $R_\star/\Gamma_0\sim N^2\times \Delta \W_\text{2D}\sim N^{1+\frac{1}{4}}$.

{\bf 3D Arrays.} For three dimensional arrays, we consider $\kk_0=k_0\uex$ in \eqnref{eq:principal_evect_array}. The emission pattern reads
\be\label{eq:mu_3D_array}
    \mu_\text{3D}(\phi,\theta) = \inv{N^2}\pare{\frac{\sin[k_0L (1-\cos\phi\sin\theta)/2]}{\sin[k_0 d(1-\cos\phi\sin\theta)/2]}}^2\pare{\frac{\sin(\frac{k_0L}{2}\sin\phi\sin\theta)}{\sin(\frac{k_0 d}{2}\sin\phi\sin\theta)}}^2\pare{\frac{\sin(\frac{k_0L}{2} \cos\theta)}{\sin(\frac{k_0d}{2}\cos\theta)}}^2.
\ee
The last term in \eqnref{eq:mu_3D_array} peaks around $\theta=\pi/2$ and its width, estimated from the distance between its first two zeros, is $\Delta\theta = 4\pi/k_0L$. For $\theta =\pi/2$, the contribution from the second term in \eqnref{eq:mu_3D_array} peaks at $\phi=0$, and its width is $\Delta\phi = 4\pi/k_0L$. Instead, the first term has a larger width $\Delta\phi = 2\sqrt{4\pi/k_0L}$. 
This means that radiation is emitted predominantly into a solid angle 
\be
\Delta\W_\text{3D} \simeq \Delta\theta\Delta\phi = \pare{\frac{4\pi}{k_0L}}^2 \sim N_\text{1D}^{-2}.
\ee
This result agrees with the calculation of Abella, Kurnit, and Hartman~\cite{S_Abella1966}. As a consequence, the scaling of the maximum emission rate for a 3D array is given by $R_\star/ \Gamma_0\sim N^2\times \Delta \W_\text{3D}\sim N^{1+\frac{1}{3}}$, as expected.

\subsection{Disordered Clouds}

The principal eigenvectors for disordered clouds of atoms are given in \eqnref{eq:Ansatz_Eigenstates} in the main text which we reprint below for convenience:
\be\label{eqS:Ansatz_Eigenstates}
	\psi_\nn(\rr) = \mathcal{N}_n 
	\left\{
	\begin{array}{ll}
	j_n(k_0r)& \quad \text{for $D=1$},\\
	J_n(k_0r)e^{\im n\phi}& \quad \text{for $D=2$},\\
	j_n(k_0 r) Y_{nm}(\theta,\phi) &\quad \text{for $D=3$},
	\end{array}
	\right .
\ee
The emission pattern from a 2D cloud is computed in the main text. In the following we consider 1D and 3D clouds.

{\bf 1D Clouds.} In the limit $k_0L\gg n^2$, the eigenstates for a one-dimensional cloud, normalized such that $\int \text{d}r \rho_\text{1D} |\psi_n(r)|^2=1$, where $\rho_\text{1D}$ is the constant density of the cloud, read
\be\label{eq:Eigenstates_Cloud_1D}
    \psi_n(r) = \sqrt{\frac{2n+1}{\pi} \frac{k_0}{\rho_\text{1D}}} \,j_n(k_0r).
\ee
Substituting \eqnref{eq:Eigenstates_Cloud_1D} and \eqnref{eq:Density_Cloud} into \eqnref{eq:Emission_Pattern_General}, and assuming the cloud to be aligned along the $\uez$ axis yields
\be\label{eq:mu_1D_clouds_def}
    \mu_{\text{1D}}^\text{cloud}(\theta) = (2n+1)\frac{k_0\rho_\text{1D}}{\pi N}\bigg\vert\int_{-L/2}^{L/2}\!\!\text{d}z\, e^{-\im k_0 z\cos\theta}j_n(k_0 z)\bigg\vert^2 = (2n+1) \pare{\frac{P_n(\cos\theta)}{\sqrt{k_0L}}}^2,
\ee
where, to compute the integral, we used the plane wave expansion $e^{-\im k_0 z \cos\theta }= \sum_{l=0}^\infty (2l+1) i^l j_l(k_0 z) P_l(\cos\theta)$ together with \eqnref{eq:Integral_two_spherical_bessel_j} in the limit $k_0L\gg 1$. As expected, the emission pattern is cylindrically symmetric with respect to the axis of the cloud. The emission pattern as a function of $\theta$ depends on the index $n$ through $P_n^2(\cos\theta)$ and it is always maximal at the edges of the cloud. We can thus interpret \eqnref{eq:mu_1D_clouds_def} as follows. Upon detecting a photon along a direction $\theta$, we can infer, as a consequence of diffraction limit, that the photon was emitted inside a solid angle $\Delta\W\sim 1/k_0L$ around that direction.

{\bf 3D Clouds.} We compute the angular distribution of a photon emitted by a 3D cloud from the collective eigenstates in \eqnref{eq:Ansatz_Eigenstates}. These eigenstates read 
\be\label{eq:Eigenstate_3DCloud_normalized}
    \psi_n(\rr) = \frac{2}{\rho_\text{3D}L^3} \spare{j_n^2(k_0L)-j_{n+1}(k_0L)j_{n-1}(k_0L)}^{-\frac{1}{2}} j_n(k_0r) Y_{nm}(\phi,\theta).
\ee
Substituting \eqnref{eq:Eigenstate_3DCloud_normalized} into \eqnref{eq:Emission_Pattern_General} and using \eqnref{eq:Density_Cloud} we obtain
\be
    \mu_\text{3D}^\text{cloud}(\phi,\theta) = \frac{3 |Y_{nm}(\phi,\theta)|^2}{2(k_0L)^2},
\ee
where we assumed $n^2\ll k_0L$.
To arrive at this result, we used the plane wave expansion and the orthonormality of the spherical harmonics.
Analogously to 1D and 2D clouds, if a photon is detected along a direction $(\phi,\theta)$, we infer that, as a consequence of diffraction limit, it was emitted in a solid angle $\Delta\W\sim 1/(k_0L)^{2}$.

\section{Scaling via SDP relaxation}\label{sec:SDP}

In Ref.~\cite{S_Mok2026} it was shown that the scaling of $R_\star$ can be obtained numerically using a semidefinite programming relaxation (SDP). Specifically, up to an additive term scaling at most as $O(N\Gamma_0)$, $R_\star$ is bounded by
\be
    \frac{1}{2\pi}R_\text{SDP} \leq R_\star \leq R_\text{SDP},
\ee
where the SDP relaxation is formulated by the convex optimization problem
\be\label{eq:R_sdp}
\begin{split}
    & R_\text{SDP} = \max_{\bold{x}\in \mathbb{R}^N} \sum_{i\neq j} \Gamma_{ij} \bold{x}_i\cdot \bold{x}_j\\
    & \text{subject to}~|\bold{x}_j|=1~\forall j=1\ldots N.
\end{split}
\ee

\begin{figure}
    \includegraphics[width=0.5\columnwidth]{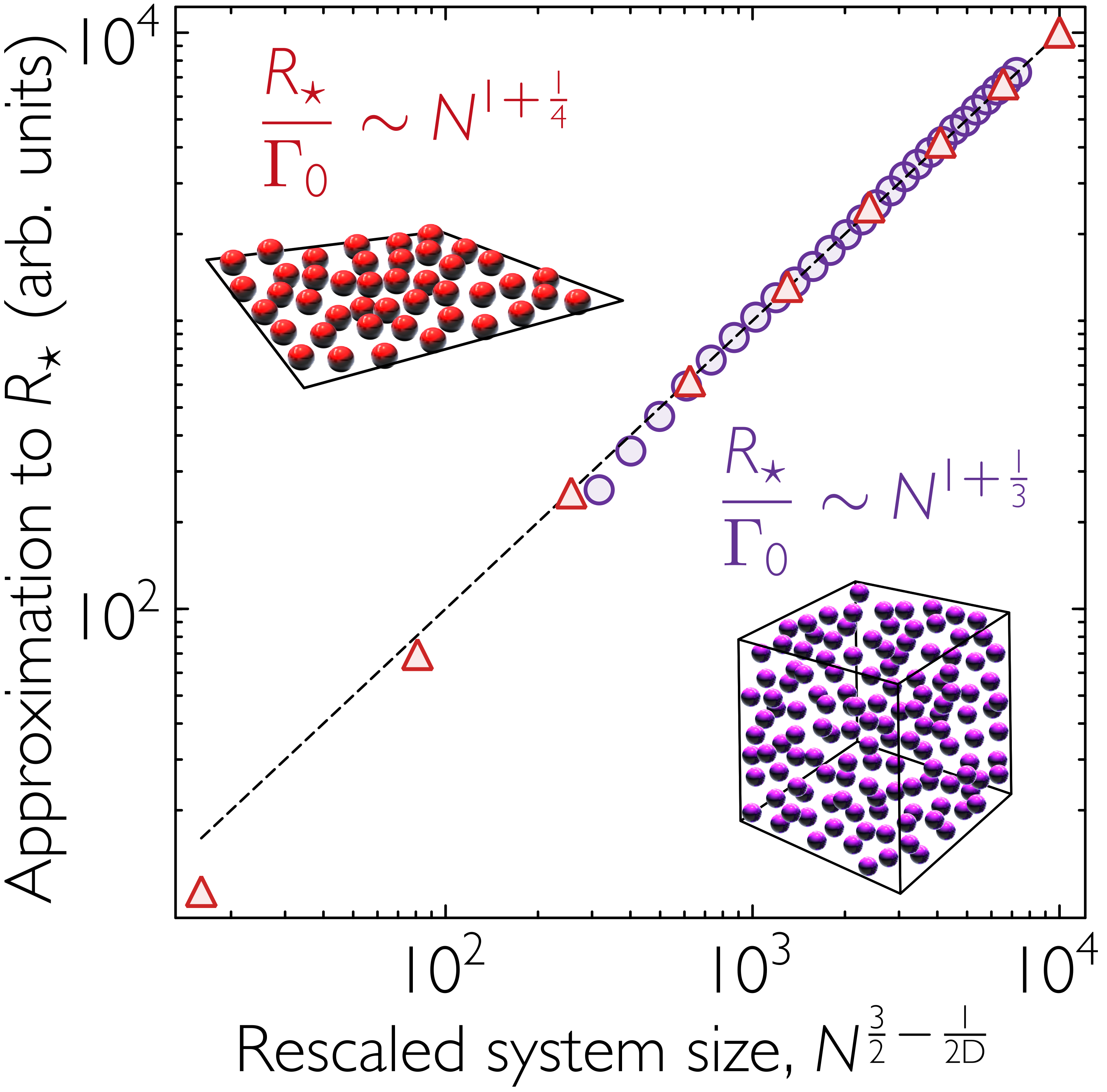}
    \caption{Scaling with system size of the numerical approximation for the maximal decay rate $R_\star$ given by the SDP solution $R_{\text{SDP}}$, for atomic clouds in 2D ({\scriptsize $\triangle$}) and 3D ($\circ$) with average interatomic separation $d/\lambda_0\simeq 0.04$ and $d/\lambda_0\simeq 0.2$ respectively. The results were averaged over 500 realizations. Error bars representing the $1\sigma$ confidence interval are too small to be visible. The dashed black line is a guide to the eye representing the dimensional scaling law $R_\star\sim N^{\frac{3}{2}-\frac{1}{2D}}\Gamma_0$. For visualization purposes, we shift the data set corresponding to each ensemble dimensionality by a multiplicative factor (a constant shift in the logarithmic scale) which does not affect the scaling.}
    \label{fig:Fig_SDP}
\end{figure}

 We solve numerically \eqnref{eq:R_sdp} using an SDP solver~\cite{S_diamond2016cvxpy} for 2D and 3D clouds, for each realization of the atomic positions. In \figref{fig:Fig_SDP} we plot the results averaged over all realizations for values of the average interatomic distance where we expect the scaling law \eqnref{eq:Scaling_Law} to hold according to \figref{fig:Fig2}. We obtain excellent agreement with the expected scaling for $R_\star$ obtained in the main text by other methods.

\section{Other electromagnetic environments: cavities and waveguides}

In the main text, we have demonstrated the validity of the scaling law $R_\star \sim N\Gamma_\text{max} \sim \Gamma_0 N\times \text{OD}$ for the case of ensembles of atoms in free space. In this section, we show that its validity extends to more general electromagnetic environments. Specifically, we study the case of atoms in cavities or along a waveguide and show that $\Gamma_\text{max}/\Gamma_0 \sim \text{OD}\sim N$. 

\subsection{Atoms in a cavity}

For atoms on resonance with a particular mode of an optical cavity---such that the cavity response can be approximated as the one of that single mode---the dissipative interaction matrix reads (see e.g. Ref.~\cite{S_AsenjoGarcia2017PRA})
\be\label{eq:Gamma_cavity}
    \Gamma_{ij}^{\text{cav}} \equiv \Gamma_\text{1D} \cos(k_c z_i)\cos(k_c z_j),
\ee
where $k_c$ is the wave vector of the cavity mode, $\Gamma_\text{1D}$ is the decay of the atom into the cavity, and $z_i$ the position of atom $i$ along the cavity axis. We assume for simplicity that the atoms are placed sufficiently close to the cavity axis, and neglect the lateral variation of the cavity mode's field.

It is immediate to see that the vector with entries
\be\label{eq:eigenvector_cavity}
    \psi_j^\text{cav} = \mathcal{N}\cos{k_c z_j}
\ee
is an eigenvector of the dissipative matrix in \eqnref{eq:Gamma_cavity} with associated eigenvalue
\be\label{eq:Gamma_cavity_eval}
    \Gamma = \pare{\Gamma_\text{1D}\sum_{j=1}^N \cos^2(k_cz_j)}= \frac{\Gamma_\text{1D}}{2} \pare{N + \sum_{j=1} ^N\frac{e^{2\im k_cz_j}+e^{-2\im k_cz_j}}{2}}, 
\ee
where the atomic position $z_j$ is uniformly distributed over the cavity length $L_c$, and the phases appearing in the complex exponentials in \eqnref{eq:Gamma_cavity_eval} are uniformly distributed on $[0,2\pi]$. 
Averaging over the uniform distribution of atomic positions, the oscillating term vanishes and we obtain $\avg{\Gamma}_\text{cnf} \sim \Gamma_\text{1D}N/2$, where $\avg{\cdot}_\text{cnf}$ denote the average over configurations. 
For a given realization, fluctuations scale as $\sim\sqrt{N}$, so $\avg{\Gamma}_\text{cnf}\sim \Gamma_\text{1D}N/2+O(\sqrt{N})$. Combined with $\norm{\boldsymbol{\psi}^\text{cav}}_1^2\sim N$, this confirms $R_\star \sim \Gamma_\text{1D}N^2$ for a gas of atoms in a single mode cavity.

\subsection{Atoms along a waveguide}

We now consider the case of atoms along a waveguide. For this scenario the dissipative matrix reads (see e.g. Ref.~\cite{S_AsenjoGarcia2017PRA})
\be\label{eq:Gamma_matrix_cavity}
    \Gamma_{ij}^{\text{wg}} \equiv \Gamma_\text{1D} \cos(k_c (z_i-z_j)) = \Gamma_\text{1D}\spare{\cos(k_c z_i)\cos(k_c z_j) + \sin(k_c z_i)\sin(k_c z_j) }.
\ee

The specific form of \eqnref{eq:Gamma_matrix_cavity} suggests to look for eigenstates of the form
\be\label{eq:eigenvector_waveguide}
    \psi_j^\text{wg} = A\cos(k_c z_j) + B\sin(k_c z_j).
\ee
Substituting this ansatz into the eigenvalue problem in \eqnref{eqS:Eigenvalue_Problem} where $\Gamma_{ij}=\Gamma_{ij}^\text{wg}$, we obtain
\be
\frac{\Gamma_\text{1D}}{2}
\begin{pmatrix}
    N + \sum_j \cos(2k_c z_j) - 2\frac{\Gamma}{\Gamma_\text{1D}} & \sum_j\sin(2k_c z_j)\\
    \sum_j\sin(2k_c z_j) & N - \sum_j \cos(2k_c z_j) - 2\frac{\Gamma}{\Gamma_\text{1D}}
\end{pmatrix}
\begin{pmatrix}
    A \\ 
    B
\end{pmatrix} = 0.
\ee
The eigenvalues are given by~\cite{S_Cardenas-Lopez2023}
\be \label{eq:Gamma_pm}
    \Gamma_\pm = \frac{\Gamma_\text{1D}}{2}\spare{N \pm \sqrt{N + \sum_{i\neq j=1}^N \cos(2k_c(z_i-z_j))}\,}.
\ee
For atoms in the mirror configuration (that is, separated by a multiple of $\lambda_c = 2\pi/k_c$), \eqnref{eq:Gamma_pm} reduces to the well known result $\Gamma_\pm = \Gamma_\text{1D}(N\pm N)/2$. For disordered ensembles, the second term inside the square root is the sum of $N(N-1)$ random variables of zero mean. From the central limit theorem, the result of the sum is zero on average and its fluctuations are of order $\sim N$. The contribution of the square root term in \eqnref{eq:Gamma_pm} thus scales at most as $\sqrt{N}$. This leads to the conclusion that $\Gamma_\pm \sim N\Gamma_\text{1D}/2$, which together with the fact that the ansatz in \eqnref{eq:eigenvector_waveguide} is delocalized, namely $|\!|\boldsymbol{\psi}^\text{wg}|\!|_1^2\sim N$, ensures the validity of the scaling law $R_\star \sim \Gamma_\text{1D}N^2$. For the case of atoms along a waveguide, it was recently shown in Ref.~\cite{S_Zhang2026} that an initially fully excited disordered ensemble of atoms dynamically achieve a peak intensity scaling as $\sim N^2$, thus saturating the upper bound $R_\star$.

\end{document}